\documentclass[article,12pt]{elsarticle}
\usepackage{lineno,hyperref}
\modulolinenumbers[5]
\usepackage{lineno}
\modulolinenumbers[5]
\usepackage{amssymb}
\usepackage{hyperref}
\usepackage{graphicx}
\usepackage{color,soul}
\usepackage{grffile}
\usepackage{booktabs}
\usepackage{times}
\usepackage{subcaption}
\usepackage{amsthm}
\usepackage{ulem}
\usepackage[titletoc]{appendix}
\usepackage{titletoc}
\usepackage{mathtools}
\usepackage{doi}
\usepackage{url}
\usepackage{tikz}
\usetikzlibrary{calc}
\usepackage{multirow}
\usetikzlibrary{patterns}
\usepackage[margin=1.0in]{geometry}
\usepackage{euscript}
\usepackage{wrapfig}
\theoremstyle{definition}

\usetikzlibrary{matrix,shapes,arrows,positioning,chains}
\usepackage{setspace}
\begin{document}
	
	%
	%

\begin{frontmatter}
		
\title{A mathematical study of the interaction between oxygen and lactate in an in-vivo and in-vitro tumor}
\author[ad]{Gopinath Sadhu \corref{mycorrespondingauthor}}
\ead{gsadhu@iitg.ac.in}
		
\author[ad]{D C Dalal}
\cortext[mycorrespondingauthor]{Corresponding author}
\address[ad]{Department of Mathematics, Indian Institute of Technology Guwahati, Guwahati, India}
\begin{abstract}
Micro-environmental acidity is a common feature of the tumor. One of the causes behind tumor acidity is lactate production by hypoxic cells of the tumor. Hypoxia is a direct result of the establishment of oxygen gradients. It is commonly observed in the tumor in an in-vitro experimental setup and also in-vivo situation. Here, we propose a mathematical model to analyses the production of lactate by hypoxic cells, and it is used as an alternative fuel by normoxic cells in tumor tissue in-vitro and in-vivo conditions. 
In this article, we study the effects of unequal oxygen concentration at the tumor boundaries on lactate status in the tumor. The effects of presence of the necrotic core in the tumor on the  lactate concentration profile is examined. The results have good agreement with experimental data and align with the theoretical findings of previous studies. The analytical results show that lactate levels are elevated in an in-vivo tumor compared to that in an in-vitro tumor. Also, during the onset of necrotic core formation, the effects of necrotic core on lactate levels are noticed.
Knowledge of the lactate status in a patient's tumor may be helpful in choosing the rightful and precious medicines for cancer treatment.
\end{abstract}
		
\begin{keyword}
	\texttt Solid tumor, oxygen, lactate, diffusion, asymmetric tumor
\end{keyword}
		
	\end{frontmatter}
	
	
\section{Introduction}
Cancer is the second leading cause of death in the world, following heart disease. In the early stages of cancer, a genetically flawed cell begins to proliferate indefinitely  in an uncontrolled manner and a mass of cells, known as a tumor is formed. The nutrients supplied by pre-existing blood vessels maintain the viability of tumor cells. As the tumor grows in size, a gradient of nutrients, particularly oxygen, is generated in the tumor core. This is due to the cells that reside near the source of oxygen consume a greater fraction of the oxygen concentration. Tumor cells require the necessary energy in a faster time compared to healthy cells. As a result, tumor cells choose different pathways to generate energy in order to maintain their uncontrollable proliferative ability. In healthy cells, energy in the form of adenosine tri-phosphate (ATP) is produced from glucose metabolism by oxidative phosphorylation (OXPHOS). Tumor cells, on the other hand, rely on aerobic glycolysis (the conversion of glucose into lactate in the presence of oxygen) for ATP production. This fact was observed by Nobel Laureate Otto Warburg a century ago  \cite{warburg1927metabolism}. The unifying feature of this altered metabolism increases glucose absorption and lactate fermentation, which ultimately aids cancer progression. Highly acidic in the tumor microenvironment is the hallmarks of cancer progression \cite{hanahan2000hallmarks,dhup2012multiple}. It promotes tumor invasion, metastasis, and therapeutic failure, particularly chemo-resistance  \cite{liu2021mechanism,simmons2017environmental}.

Mathematical models aid in understanding the underlying mechanism of biological phenomena \cite{sadhu2023,yadav2023multiscale}. \citet{2013webb} investigated the role of lactate and H+ ions in regulating intracellular pH using a dynamical system. The same group also investigated the effects of oxygen and pH on tumor growth and morphology \cite{al2014cellular}. They observed that oxygen and extracellular pH influence tumor cellular growth and phenotype transformation. \citet{gatenby2006acid} formulated an acid-mediated tumor invasion continuum model which provides a simple mechanism for coupling altered glucose metabolism with tumor cells' ability to produce invasive malignancies. Acidity also influences the tumor cell-cell cycle and the development of necrotic core \cite{tindall2012modelling}. As a result, micro-environmental acidity significantly impacts progression and plays a role in therapeutic failure.

For glucose metabolism, tumor cells in the oxygen-rich region depend primarily on OXPHOS, whereas cells in the hypoxic region of the tumor depend on anaerobic phosphorylation and generate lactate abundantly. As a result, hypoxic region is the primary source of lactate in tumor tissue \cite{reshkin2014role}. The oxygen concentration in the tissue is high in the vicinity of the oxygen source site (e.g., the local pre-existing blood vessel in the case of an avascular tumor) and gradually decreases with the diffusional distance from the oxygen source. Many mathematical models in existing literature tend to overlook crucial and realistic scenarios when it comes to the dynamics of tumor growth. These models generally make the simplifying assumption that tumor tissue uniformly maintains a constant oxygen concentration at its boundaries. Consequently, they have explored the intricate interplay of acidosis within tumors and its impact on both tumor growth and the generation of metabolic waste \cite{casciari1992mathematical,tindall2012modelling,Byrne2002,ward1997mathematical}. \citet{fiandaca2021mathematical} recently developed a mathematical model to study the effects of acidosis on tumor growth under unequal oxygen concentrations at the tumor boundaries and they assumed a lactate concentration threshold above which cell death occurs. However, as tumor biology advances, scientists have discovered that lactate is more than just a byproduct of glucose synthesis, and the role of lactate is not limited to healthy tissue damage. Instead, lactate acts as a metabolic fuel for well-oxygenated tumor cells or a signaling molecule for spreading network, as discussed in the articles \cite{IPPOLITO2019153,baltazar2020lactate}. According to \citet{sonveaux2008}, lactate is a major substrate that fuels the oxidative metabolism of oxygenated tumor cells. 
 
In this article, we have proposed a one-dimensional mathematical model to understand the lactate and oxygen interplay under the assumption that lactate acts as an alternative fuel in the oxygen-rich region of a tumor for both in-vivo and in-vitro situations. This model is solved analytically. The presented model has the ability to compute the sizes of both pre-necrotic and fully developed tumors. Also, in the current study, the lactate and oxygen dynamics with various size of tumor are explored  qualitatively. The effects of the presence of the necrotic core in tumor on lactate accumulation is examined. Our results are consistent with experimental observations. Lactate plays a  crucial role in tumor progression, invasion and disease prognosis. Our results may help clinicians to plan therapies and pharmacologists to design drugs for maximum benefits.

\section{The formulation of problem}
\subsection{Physical background}
Oxygen is a pivotal component of cell respiration. Oxygen diffuses into the tissue domain from blood vessels. In case of an avascular tumor, the oxygen demand is fulfilled by the nearby blood vessels. As the tumor grows, oxygen shortage is sensed in the region away from blood vessels. In that region, tumors have a threshold value of oxygen concentration $C_H$; below that, tumor cells lose their proliferation ability but they remain viable. This region is known as hypoxic region (i.e., $C\le C_H$, where $C$ represents oxygen concentration). For large tumors, oxygen concentration reaches a threshold value $C_N$; as a results, tumor cells become dead. In a region, where $C\le C_N$ is maintained, is known as the necrotic region.
So, it can be assumed that oxygen is consumed by tumor cells that are present in a viable state. In the hypoxic region, glucose metabolism of tumor cells shifts toward anaerobic metabolism, producing lactate as the main byproduct. So, lactate can be considered as a pivotal contributor to the acidification of the tumor ecosystem. Tumor cells in well-oxygenated regions uptake that lactate as an alternative metabolic fuel.

\subsection{Governing equations}
	
{\bf Oxygen diffusion equation:}
The steady-state oxygen transport is given as \cite{lewin2020three}
\begin{equation}
	0=D_c\frac{d^2 C}{d x^2} -\gamma_c C H(C-C_N),
	\label{eq:Oxygen_diffusion}
\end{equation}
where $C$ is the oxygen concentration, $D_c $ is the diffusion coefficient and $\gamma_c$ is the rate at which the tumor cells consume oxygen. Last term of the right hand side (RHS) of Eq. \eqref{eq:Oxygen_diffusion} interprets that oxygen is only consumed by viable cells and  $H(x)$ is the Heaviside function and defined as
\begin{align}
	H(x)=
	\begin{cases}
		&1,\; x>0,\nonumber\\
		&0,\; \text{otherwise}.
	\end{cases}
\end{align}

{\bf Lactate diffusion equation:} The steady-state lactate transport is given as
\begin{equation}
	0=D_l\frac{d^2 L}{d x^2} +k LH(C-C_N)H(C_H-C)-\gamma_l LH(C-C_H)+d H(C-C_N),
	\label{eq:pH_diffusion}
\end{equation}
where $L$ is the lactate concentration, $D_l $ is the diffusion coefficient. The second term on the RHS represents that lactate acid is produced by tumor cells in hypoxic region (i.e, $C_N\le C \le C_H$) only with the rate $k$. The third term on the RHS of Eq. \eqref{eq:pH_diffusion} denotes that lactate is consumed by the tumor cells in the rich-oxygen area (i.e., $C\ge C_H$) of the tumor with the rate $\gamma_l$. Lactate can be produced even in aerobic conditions \cite{brooks1986lactate} and also through other means by live cells. The last term in the equation, represented by $d$, accommodates this observation.
\subsection{Boundary conditions}
The boundary conditions of the model equations are guided by in-vitro and in-vivo circumstances. 

{\bfseries In-vivo case:} When the tumor is present in an in-vivo environment, nutrient availability is higher near the source than  that away from the source. So, oxygen concentrations are unequal at boundaries. Lactate generated from glycolysis has a higher chance of being flushed out at the oxygen source edge as this edge is very near to the blood vessel. Lactate accumulation starts far from the oxygen source site. Hence, ``no flux" boundary condition is imposed for lactate at the that boundary \cite{fiandaca2021mathematical}.
\begin{align}
	C(0)&=C_l,  C(R)=C_r\;\; \text{with}\; C_l>C_r
	\label{eq:invivo_oxy_bdd}\\
	L(0)&=L^{*}, \; \frac{d L}{d x}=0\;\; \text{at}\;\; x=R,
	\label{eq:invivo_lactate_bdd}
\end{align}
where $C_l$, $C_r$  are oxygen concentrations at left and right boundaries and  $L^{*}$ is the lactate concentration at blood vessel.

{\bfseries In-vitro case:} When a tumor presents in an in-vitro environment, the availability of nutrients (in the present study, oxygen) is limitless at both the boundaries. The metabolic waste, like lactate, is flushed out through the edges. For the sake of simplicity, lactate is flash out in similar way as in in-vivo situation. Hence, the reasonable boundary conditions are considered as,
\begin{align}
	C(0)=C(R)=C_l,
	\label{eq:invitro_oxy_bdd}\\
	L(0)=L(R)=L^{*}.
	\label{eq:invitro_lactate_bdd}
\end{align}
\section{Non-dimensionalized model}
Let $R$ be the length of the tissue, $\hat{C}$ be the characteristic oxygen concentration, 
 and $L_0$ be the characteristic lactate concentration which is same as normal physiologic value of lactate in blood. 
So, the dimensionless variables are given as, 
\begin{equation}
		x^{'}=\frac{x}{R}, ~C^{'}=\frac{C}{\hat{C} }  \text{ and }  ~L^{'}=\frac{L}{L_{0} } 
		\label{eq:non_dim_para}
\end{equation}
The parameters and variables are made dimensionless in the following way:  $C_H=C_H^{'}\hat{C}$,  $C_N=C_N^{'}\hat{C}$, $d^{'}=\frac{dD_c}{\gamma_cD_lL_0}$, $C_l^{'}=\frac{C_l}{\hat{C}}$, $C_r^{'}=\frac{C_r}{\hat{C}}$ and $\phi=R\sqrt{\frac{\gamma_c}{D_c}}$, where $\phi$ is the Thiele modulus. For the sake of simplicity, the `prime' symbol is dropped over dependent and independent variables from the governing equations.

The oxygen diffusion Eq. \eqref{eq:Oxygen_diffusion} in dimensionless form becomes,
\begin{equation}
0=\frac{d^2 C}{d x^2} -\phi^2 C H(C-C_N).
	\label{eq:Non_dim_oxy}
\end{equation}
The non-dimensional form of oxygen concentrations at both the boundaries is given as,
\begin{eqnarray}
		C=C_{l}\;\; \mbox{at} \;\; x=0,\nonumber \\
		C=C_{r}\;\;\; \mbox{at} \;\; x=1.
		\label{eq:non_dim_oxygen_boundary}
\end{eqnarray}  
It should be noted that $\hat{C}=C_l$ in the current study. 

The non-dimensional form of lactate diffusion equation is given as,
\begin{equation}
	0=\frac{d^2 L}{d x^2} +\phi^2 \alpha LH(C-C_N)H(C_H-C)-\phi^2\beta LH(C-C_H)+d\phi^2H(C-C_N),
	\label{eq:non_dim_lactate}
\end{equation}
where $\alpha=\frac{D_ck}{D_l\gamma_c}$, $\beta=\frac{D_c\gamma_l}{D_l\gamma_c}$ are two non-dimensional numbers. Here, it should be noted that $L_0=L^{*}$. Hence, the corresponding boundary conditions are given as
\begin{align}
	L(x=0)=1,\;\; \frac{d L}{d x}=&0\; \text{at}\; x=1\;\;\vspace{2cm} \text{ (In-vivo case)},\\
	L(x=0)=L(x=1)=&1 \;\; \text{(In-vitro case)}.
\end{align}

\begin{table}[h!]
	\centering
	\caption{Parameters values.}
	\begin{tabular}{ p{2cm} p{4.5cm} p{4cm} p{2cm}}
		\hline
		
		Parameter &Physical description& Dimensional value& Reference\\
		\hline
		$D_c$    &Oxygen diffusivity &$1820\; \mu m^2s^{-1}$ & \cite{oxy_diff_coeff} \\
		$D_l$    &Lactate diffusivity &$1100\; \mu m^2s^{-1}$ & \cite{lactate_D_coeff} \\
		$C_H$  &Hypoxic threshold value&$5\; mmHg$ & \cite{1997acid_oxy}\\
		$C_N$  &Necrotic threshold value&$0.2-0.5\; mmHg$ & \cite{1997acid_oxy}\\
		$\gamma_c$ &Oxygen consumption rate&$0.5685\; s^{-1}$  & Estimated \\
		$\gamma_l$&Lactate uptake rate    &$1.3715\;s^{-1}$  &Estimated\\ 
		$k$& Rate of lactate production in anaerobic glycolysis &$0.06322\; s^{-1}$ & Estimated \\
		$d$& Lactate produce by other than anaerobic glycolysis &$0.015\; mMs^{-1}$ & \cite{phypers2006lactate}\\ 
		$C_l$    &Oxygen concentration near the blood vessel&$1.43\times 10^{-2} mM$& \cite{1997acid_oxy} \\
		$L_0$ & Lactate in blood vessel in normal physiologic condition & $2\; mM$ & \cite{1104501691}\\
		\hline
		\end{tabular}
	\label{tab:parameter}
\end{table}

\section{Analytical Solution}
The governing equations are solved with the help of analytical methods. Typically, when a tumor reaches a size beyond a few millimeters, it undergoes spatial differentiation into three distinct zones: a normoxic region characterized by sufficient oxygen levels ($C\ge C_H$), a hypoxic region with moderate oxygen deficiency ($C_N<C<C_H$), and a necrotic region experiencing severe hypoxia ($C\le C_N$). At this stage, the tumor is considered fully developed. Assuming that $\phi$ i.e., length of tumor tissue $R$ is large enough that the tumor is fully developed.
\begin{figure}[h!]
	\begin{subfigure}{0.49\textwidth}
		\centering
		\includegraphics[scale=0.5]{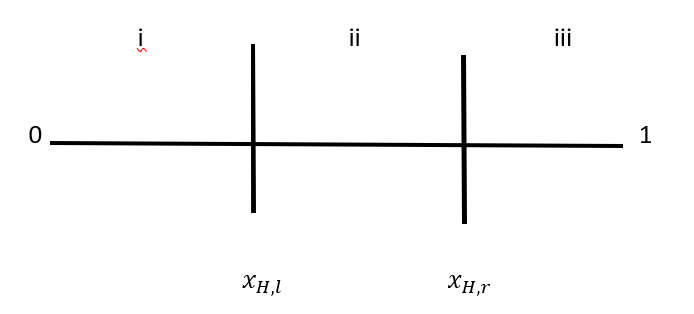}
		\caption{}
		\label{fig:pre_necro} 
	\end{subfigure}	
	\begin{subfigure}{0.49\textwidth}
		\centering
		\includegraphics[scale=0.5]{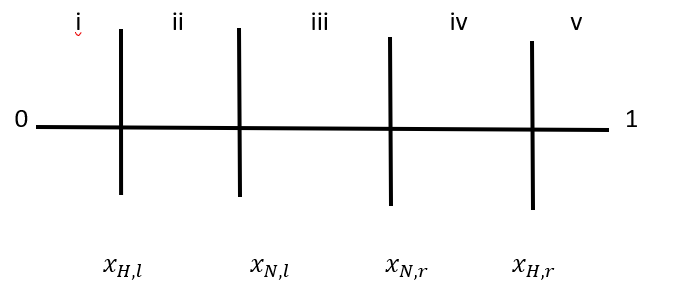}
		\caption{}
		\label{fig:full_tumor} 
	\end{subfigure}	
	
	\caption[Two numerical solutions]{The schematic region of a tumor (a) having only hypoxic region, not necrotic core at center and (b) tumor have necrotic core and hypoxic region (fully developed tumor).}
	\label{fig:scheme_tumor}
\end{figure}
Suppose $x_{H,l}$ and $x_{H,r}$ are the interface points of hypoxic region and normoxic region, and $x_{N,l}$ and $x_{N,r}$ are the interface points of necrotic region and hypoxic region, where $l$ denotes the left side and $r$ denotes the right side point in Figure \ref{fig:pre_necro}-\ref{fig:full_tumor}. 
The positions of these interface points (i.e., $x_{N,l}, x_{N,r}, x_{H,l}, x_{H,r}$) remain undetermined. It is assumed that oxygen is consumed by live tumor cells. The oxygen diffusion equation \eqref{eq:Non_dim_oxy} for a fully developed tumor can be rewritten as,
\begin{equation}\label{eq:oxy_no_necro}
	\left.\begin{aligned}
		0&=\frac{d^2C}{d x^2}-\phi^2 C, \;\; x\in[0,x_{N,l}]\cup[x_{N,r},1],\\
		0&=\frac{d^2C}{d x^2},\;\; x\in[x_{N,l},x_{N,r}].
	\end{aligned}\right\}
\end{equation}
Using boundaries conditions $C(0)=C_l$, $C(1)=C_r$, $C(x_{N,l})=C_N$ and $C(x_{N,r})=C_N$, the solution of Eqs. \eqref{eq:oxy_no_necro} are given as,
\begin{equation}\label{eq:solu_oxygen}
C(x)= 
\begin{cases}
	\frac{1}{\sinh(\phi x_{N,l})}\left(C_l \sinh(\phi(x_{N,l}-x))+C_N \sinh(\phi x)\right), &0\leq x\leq x_{N,l} \\
	C_N,& x_{N,l}\leq x\leq x_{N,r} \\
	\frac{1}{\sinh(\phi(1-x_{N,r}))}\left(C_r \sinh(\phi(x-x_{N,r}))+C_N \sinh(\phi (1-x))\right) & x_{N,r}\leq x\leq 1.
\end{cases} 
\end{equation} 
At the interface points $x_{N,l}, x_{N,r}$, it is assumed that $C$ has continuous fluxes. Using this criteria in Eq. \ref{eq:solu_oxygen}, we have
\begin{align}
	x_{N,l}&=\frac{1}{\phi}\ln\left(\frac{C_l+\sqrt{C_l^2-C_N^2}}{C_N}\right),
	\label{eq:left_necro}\\
	x_{N,r}&=1-\frac{1}{\phi}\ln\left(\frac{C_r-\sqrt{C_r^2-C_N^2}}{C_N}\right).
	\label{eq:right_necro}
\end{align}
The question remains that what is the maximum value of $\phi$ for which tumor does not have any necrotic region i.e., oxygen concentration value attains $C_N$ at a point in the tumor region. That means $C$ has a minimum value $C_N$ at a single point in $[0,1]$.
	
Prior to the formation of necrotic region in the tumor, the corresponding oxygen concentration is given as,
\begin{equation}\label{eq:no_necro}
C(x)=\frac{1}{\sinh(\phi)}\left(C_r\sinh(\phi x)+C_l\sinh(\phi(1-x))\right)\; \text{in}\;\; x\in[0,1].
\end{equation}
$C$ in Eq. \eqref{eq:no_necro} has minimum value at
\begin{equation}\label{eq:x_min}
x_{min}=\frac{1}{2\phi}\ln\left(\frac{\frac{C_l}{C_r}e^{\phi}-1}{1-\frac{C_l}{C_r}e^{-\phi}}\right). 
\end{equation} 
The minimum value of $C$ can not exceed $C_N$. By solving $C(x_{min})=C_N$, one can get the corresponding value of $\phi$ (says it is $\phi_{max}$).

For an in-vivo case ($C_l\neq C_r$), $C(x_{min})=C_N$ becomes a transcendental equation of $\phi$, which is given as,
\begin{align}
	AC_N^2\cosh(3\phi)-(4C_l^2+C_N^2(1+A^2))\cosh(2\phi)+(8C_l(C_r+2AC_l)-AC_N^2)\cosh(\phi)=\nonumber\\4C_l^2+2(C_r+2AC_l)^2-C_N^2(1+A^2),
	\label{eq:transcendental}
\end{align}
where $A=\dfrac{C_l}{C_r}$. This equation is solved by Newton-Rapshon method to obtain the approximate values of $\phi_{max}$.

For in-vitro case  ($C_l=C_r$), $x_{min}=\frac{1}{2}$ (from Eq. \ref{eq:x_min}) and the corresponding $\phi_{max}=2\cosh^{-1}(\dfrac{C_l}{C_N})$ (from Eq. \eqref{eq:transcendental}).


From these results, one can find that the maximum length $R_{max}$ (from dimensionless Thiele modulus $\phi_{max}$) of the tumor before formation of the necrotic core for both in-vivo and in-vitro cases if $C_N, C_l \; \text{and} \; C_r$ are explicitly known.

The interface points $x_{H,l}\;\text{and}\;x_{H,r}$ between normoxic and hypoxic regions can be easily obtained from Eqs. \eqref{eq:solu_oxygen} and \eqref{eq:no_necro} by putting $C=C_H$ for a tumor before and after the formation of the necrotic core in an in-vivo and in-vitro situations.
 
 {\bfseries Lactate dynamics:} Here, lactate dynamics is examined within a tumor tissue for two situations (i) pre-necrotic and (ii) post-necrotic under in-vivo and in-vitro scenarios.
 
 In the case (i), before tumor develops necrotic core (i.e., for $\phi\le \phi_{max}$), tumor tissue can be categorized into three parts. The central part is occupied by hypoxic cells and it is surrounded by normoxic cells (Figure \ref{fig:pre_necro}).
 Hence, the governing equation \eqref{eq:non_dim_lactate} takes the form for this case as
 \begin{equation}\label{eq:lactate_pre_necro}
 	\left.\begin{aligned}
 		0&=\frac{d^2 L}{d x^2}-\phi^2\beta L+d\phi^2, \;\;\text{in}\;\; x\in[0,x_{H,l}]\cup[x_{H,r},1]\\
 		0&=\frac{d^2 L}{d x^2}+\phi^2\alpha L+d\phi^2 \;\;\text{in}\;\; x\in[x_{H,l},x_{H,r}].
 	\end{aligned}\right\}
 \end{equation}
Its corresponding solutions are given as,
 \begin{equation}\label{eq:solu_lactate}
 	L(x)= 
 	\begin{cases}
 		a_{1}\cosh(\phi\sqrt{\beta}x)+a_2\sinh(\phi\sqrt{\beta}x)+\frac{d}{\beta}\;\; &\text{in}\;\;x\in[0,x_{H,l}],\\
 	a_{3}\cos(\phi\sqrt{\alpha}x)+a_4\sin(\phi\sqrt{\alpha}x)-\frac{d}{\alpha}\;\; &\text{in}\;\; x\in[x_{H,l},x_{H,r}],\\
 	a_{5}\cosh(\phi\sqrt{\beta}x)+a_{6}\sinh(\phi\sqrt{\beta}x)+\frac{d}{\beta}\;\; &\text{in}\;\;x\in[x_{H,r},1].
 	\end{cases} 
 \end{equation} 
 The values of the coefficients $a_i$ for $i=1,\dots,6$ are determined by ensuring the continuity of $L$ and its fluxes at the interface points.
 
For the case (ii), when tumor is fully developed i.e., tumor has necrotic core at center and proliferative rim at the outer side and in between tumor has hypoxic cells ($\phi>\phi_{max}$) and  lactate is generated by anerobic glycolysis in regions ii,iv and it is consumed in regions i,v and not consumed and produced in region iii in Figure \ref{fig:full_tumor}. The lactate diffusion Eq. \eqref{eq:non_dim_lactate} is rewritten as
\begin{equation}\label{eq:lac_full_tumor}
	\left.\begin{aligned}
		0&=\frac{d^2 L}{d x^2}-\phi^2\beta L+d\phi^2 \;\;\text{in}\;\; x\in[0,x_{H,l}]\cup[x_{H,r},1],\\
		0&=\frac{d^2 L}{d x^2}+\phi^2\alpha L+d \phi^2 \;\;\text{in}\;\; x\in[x_{H,l},x_{N,l}]\cup[x_{N,r},x_{H,r}],\\
		0&=\frac{d^2 L}{d x^2} \;\; \text{in}\;\; x\in [x_{N,l},x_{N,r}].
	\end{aligned}\right\} 
\end{equation}
 The solutions of the Eqs. \eqref{eq:lac_full_tumor} are given by,
 \begin{equation}\label{eq:solu_lactate1}
 	L(x)= 
 	\begin{cases}
 		a_{1}\cosh(\phi\sqrt{\beta}x)+a_2\sinh(\phi\sqrt{\beta}x)+\frac{d}{\beta},\;\; &\text{in}\;\;x\in[0,x_{H,l}]\\
 		a_{3}\cos(\phi\sqrt{\alpha}x)+a_4\sin(\phi\sqrt{\alpha}x)-\frac{d}{\alpha},\;\; &\text{in}\;\; x\in[x_{H,l},x_{N,l}]\\
 		a_5x+a_6,\;\; &\text{in}\;\; x\in [x_{N,l},x_{N,r}]\\
 		a_{7}\cos(\phi\sqrt{\alpha}x)+a_8\sin(\phi\sqrt{\alpha}x)-\frac{d}{\alpha},\;\; &\text{in}\;\; x\in[x_{N,r},x_{H,r}]\\
 		a_{9}\cosh(\phi\sqrt{\beta}x)+a_{10}\sinh(\phi\sqrt{\beta}x)+\frac{d}{\beta},\;\; &\text{in}\;\;x\in[x_{H,r},1].
 	\end{cases}
\end{equation}
At the interface points i.e., at $x=x_{H,l}, x_{H,r}, x_{N,l}, x_{N,r}$, $L$ and its flux terms are continuous. Using this virtue and the boundary conditions (\eqref{eq:invivo_lactate_bdd}, \eqref{eq:invitro_lactate_bdd}), the coefficients $a_i$'s (for $i=1,2,\dots,10$) can be evaluated.
\section{Results}
\subsection{Validation with experimental observations and parameter estimation}
The parameters involved in the present model are not available in a single existing experimental study. Oxygen diffusivity, lactate diffusivity, consumption rates of nutrients and production rate of lactate may not be always constant rather these are dependent on cellular composition, temperature, tumor type, tumor location and time. For the sake of simplicity, we assume that these parameters are constant throughout the study and their values are taken form the exiting literature listed in Table \refeq{tab:parameter}. Those parameters are not available according to present model requirement, these parameters are estimated. Oxygen partial pressure near the blood vessel in in-vivo situation remains constant and it equals to $13.5\; mmHg$ \cite{1997acid_oxy}. According to Henry's law, oxygen partial pressure relates with oxygen concentration and it's solubility coefficient can be expressed by the relation, oxygen concentration $=$ oxygen partial pressure $\times$ solubility coefficient (here, solubility coefficient is $1.05\times10^{-3}\; mM/mmHg$). This gives an estimate of $C_l$ as $1.43\times 10^{-2} \; mM$. The oxygen consumption rate of tumor cells is estimated by the least square fitting to the spatial data of \citet{1997acid_oxy}. It gives Thiele modulus $\phi=7.07$ and non-dimensional necrotic threshold value as $0.03488$. It follows that $\gamma_c=0.5685\; s^{-1}$ with $C_N=0.469\; mmHg$. The obtained $C_N$ is consistent with the experimental data \cite{1997acid_oxy}. 
\begin{figure}[h!]
	\centering
	\includegraphics[scale=0.5]{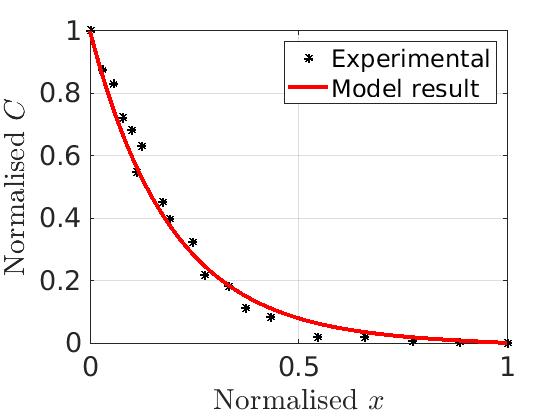}
	\caption{Comparison of present simulation results with the results of  \citet{1997acid_oxy}.}
\end{figure}

\begin{table}[h!]
	\centering
	\caption{Non dimensional parameters value are obtained using Table \ref{tab:parameter}.}
	\begin{tabular}{ p{2cm} p{2.5cm}}
		\hline
		
		Parameter & value\\
		\hline
		$C_H$  &$0.37-0.74$\\
		$C_N$  &$0.0348$ \\
		$C_l$   &$1$\\
		$\alpha$ &$0.184$ \\
		$\beta$&$3.42$\\ 
		$d$& $0.021$\\ 
		$L^{*}$ & $1$ \\
		\hline
	\end{tabular}
	\label{tab:Non_dim_parameter}
\end{table}
The non-dimensional values of parameters as shown in Table \ref{tab:Non_dim_parameter} are utilized for subsequent analysis. It may be noted that oxygen partial pressure in an in-vitro situation is much higher than that in an in-vivo situation. In an in-vitro case, $C_l=C_r$ is maintained always. However, in an in-vivo situation, $C_r$ and $C_l$ are not equal. Hence, $C_r$ is treated as a model parameter in the current study to  visualize the effects of oxygen gradient on lactate dynamics. 
\subsection{Oxygen profiles of a tumor prior to the formation of a necrotic core in in-vitro and in-vivo settings} 
\begin{figure}[h!]
	\begin{subfigure}{0.49\textwidth}
		\centering
		\includegraphics[scale=0.08]{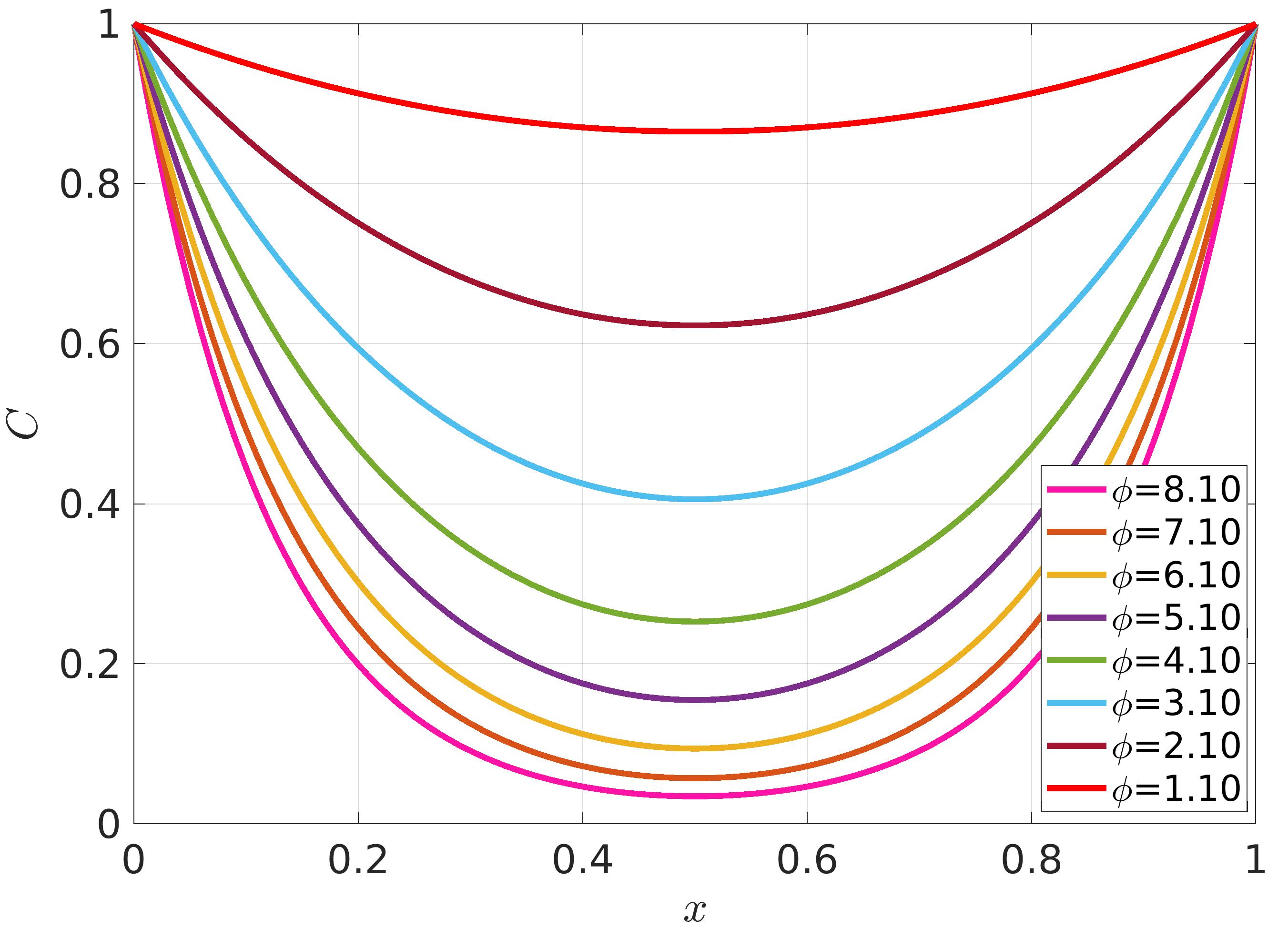}
		\caption{}
		\label{fig:oxygen} 
	\end{subfigure}	
	\begin{subfigure}{0.49\textwidth}
			\centering
			\includegraphics[scale=0.08]{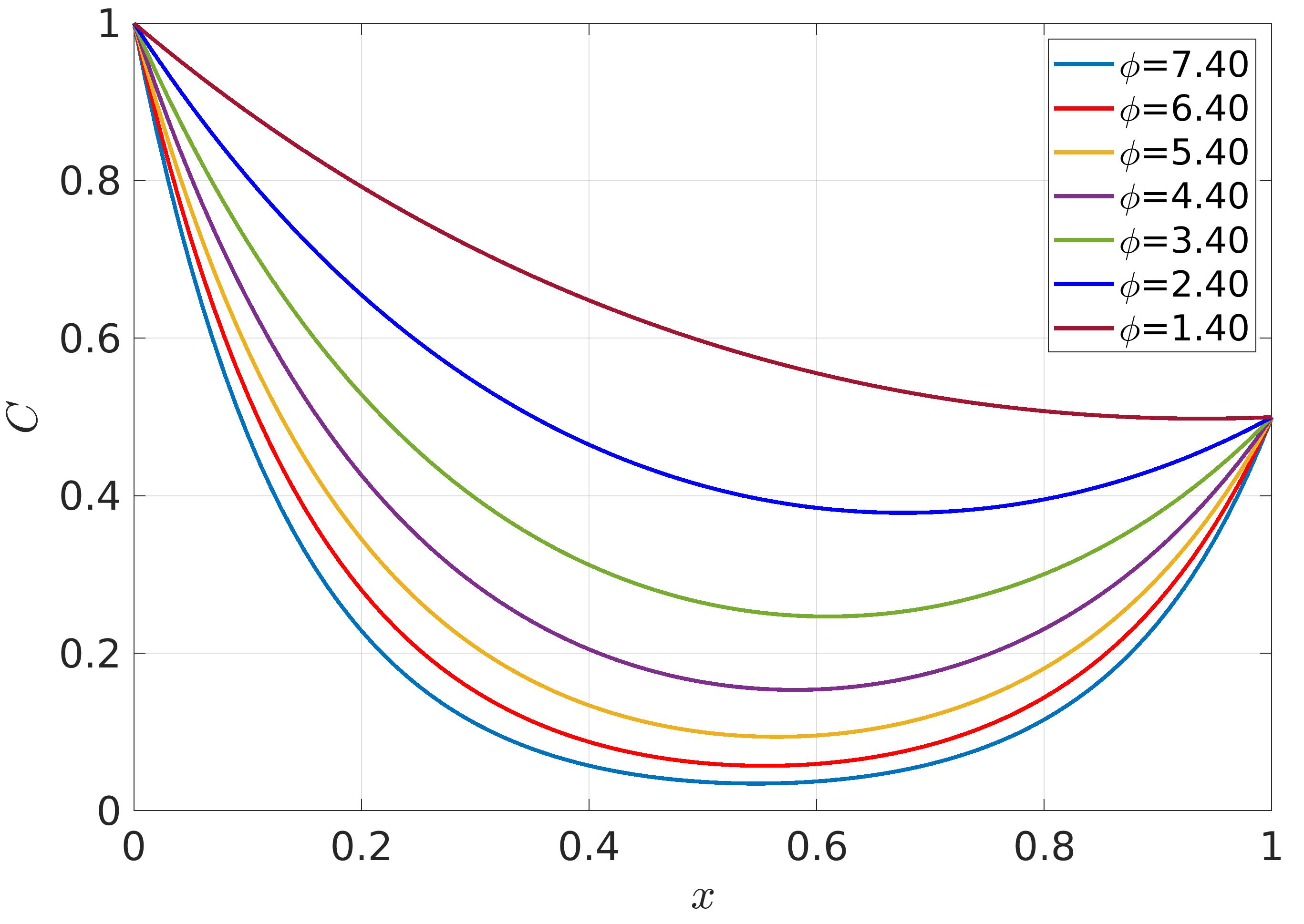}
			\caption{}
			\label{fig:oxygen05} 
		\end{subfigure}	
	\begin{subfigure}{0.49\textwidth}
			\centering
			\includegraphics[scale=0.08]{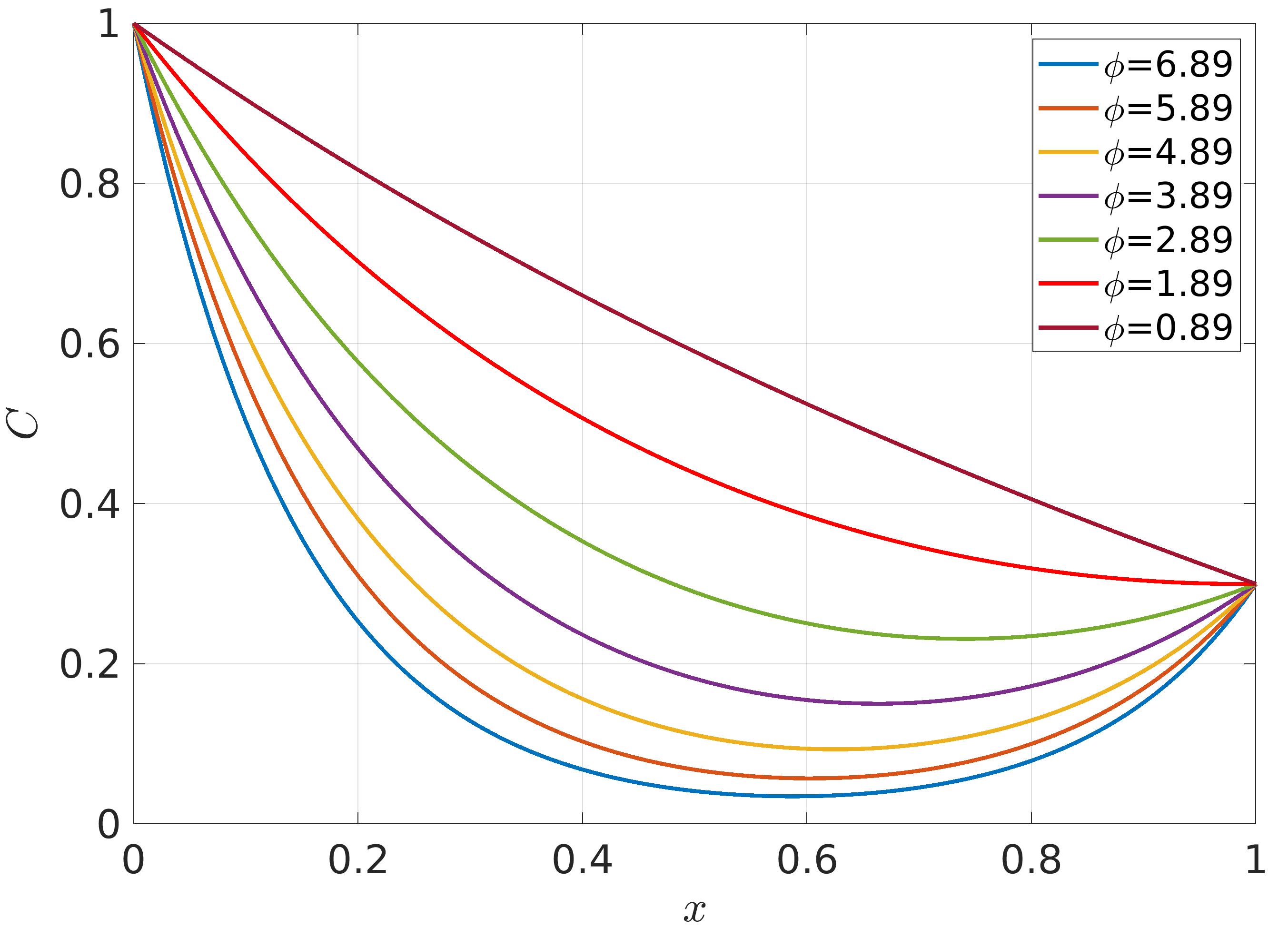}
			\caption{}
			\label{fig:oxygen03}
		\end{subfigure}
	\begin{subfigure}{0.49\textwidth}
		\centering
		\includegraphics[scale=0.08]{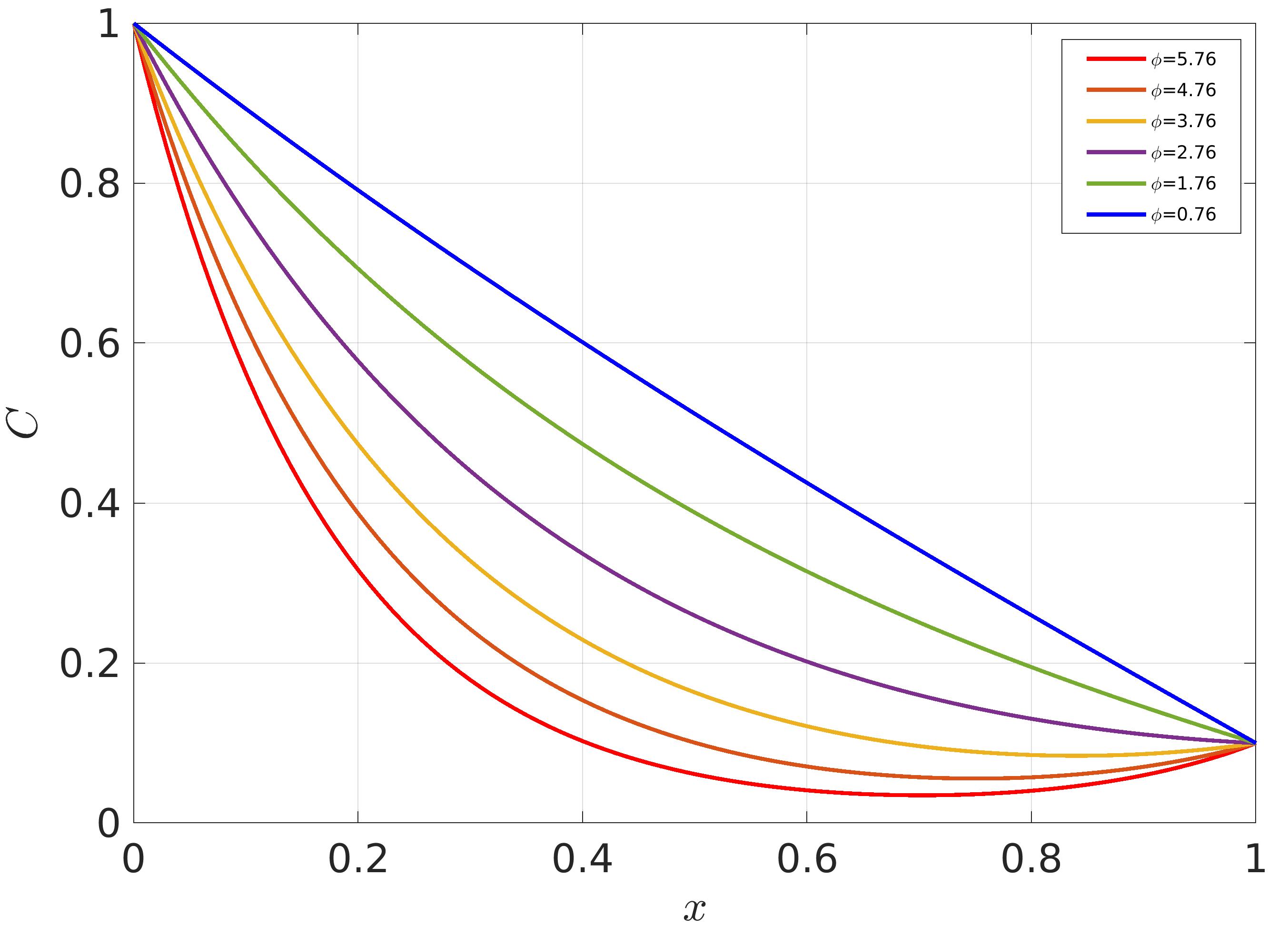}
		\caption{}
		\label{fig:oxygen01}
	\end{subfigure}
	
	\caption[Two numerical solutions]{Oxygen concentration profile in pre-necrotic tumor for various values of $\phi\; (\le\phi_{max})$ with (a) $C_l=1$, $C_r=1$ and $\phi_{max}=8.10$,  (b) $C_l=1$, $C_r=0.5$ and $\phi_{max}=7.40 $, (c) $C_{l}=1$, $C_{r}=0.3$ and $\phi_{max}=6.89 $ and (d) $C_{l}=1$, $C_{r}=0.1$ and $\phi_{max}=5.77$.}
	\label{fig:oxygen_tumor}
\end{figure}

The tumor tissues' oxygen profiles prior to development of a necrotic core are depicted in Figure \ref{fig:oxygen_tumor}. One can see that oxygen concentration falls with the increase in value of $\phi$ (i.e., tumor tissue length $R$). So, as the size of the tumor increases, the oxygen concentration starts to decrease inside the tumor. It happens because oxygen is consumed to maintain livelihood of tumor cells and mitotic cell division. It is observed that oxygen is distributed symmetrically about the center point $x=0.5$ (Figure \ref{fig:oxygen}). At this point, the oxygen concentration inside the tumor reaches the necrotic threshold value $C_N$ for in-vitro situation. However, for an in-vivo situation, oxygen concentrations at the boundaries are not equal, and the corresponding oxygen profiles are displayed in Figures \ref{fig:oxygen05},\ref{fig:oxygen03}, \ref{fig:oxygen01}. In these cases, one can observe that oxygen concentration profiles are no longer symmetric about $x=0.5$. For an in-vivo and in-vitro setting, the tumor with entire viable cells, the corresponding Thiele Modulus ($\phi_{max}$ from Eq. \eqref{eq:transcendental}) and the points ($x_{min}$ from Eq. \eqref{eq:x_min}) where oxygen concentration attains the necrotic threshold value ($C_N$) are shown in the Table \ref{tab:Theile number}. 

\begin{table}[h!]
	\centering
	\caption{Thiele Modulus and the point where oxygen touches $C_N$  .}
	\begin{tabular}{ p{5cm} p{5cm} p{2cm} }
		\hline
		
		Condition &Thiele Modulus ($\phi_{max}$)& $x_{min}$\\
		\hline
		$C_l=C_r=1$    &$8.10$ &$0.5$   \\
		$C_l=1, C_r=0.5$ &$7.40$ &$0.5468$\\
		$C_l=1, C_r=0.3$  &$6.89$ &$0.5875$\\
		$C_l=1, C_r=0.1$  &$5.77$ &$0.70$\\
		\hline
	\end{tabular}
	\label{tab:Theile number}
\end{table}
From Table \ref{tab:Theile number}, one can conclude that the maximum size of the pre-necrotic tumor in an in-vitro situation is larger than that in an in-vivo situation. The point at which the oxygen concentration touches the necrotic threshold value shifts toward a lower oxygen edge.
\subsection{Oxygen profiles after the formation of a necrotic core in tumor in in-vitro and in-vivo scenarios} 
\begin{figure}[h!]
	\begin{subfigure}{0.49\textwidth}
		\centering
		\includegraphics[scale=0.08]{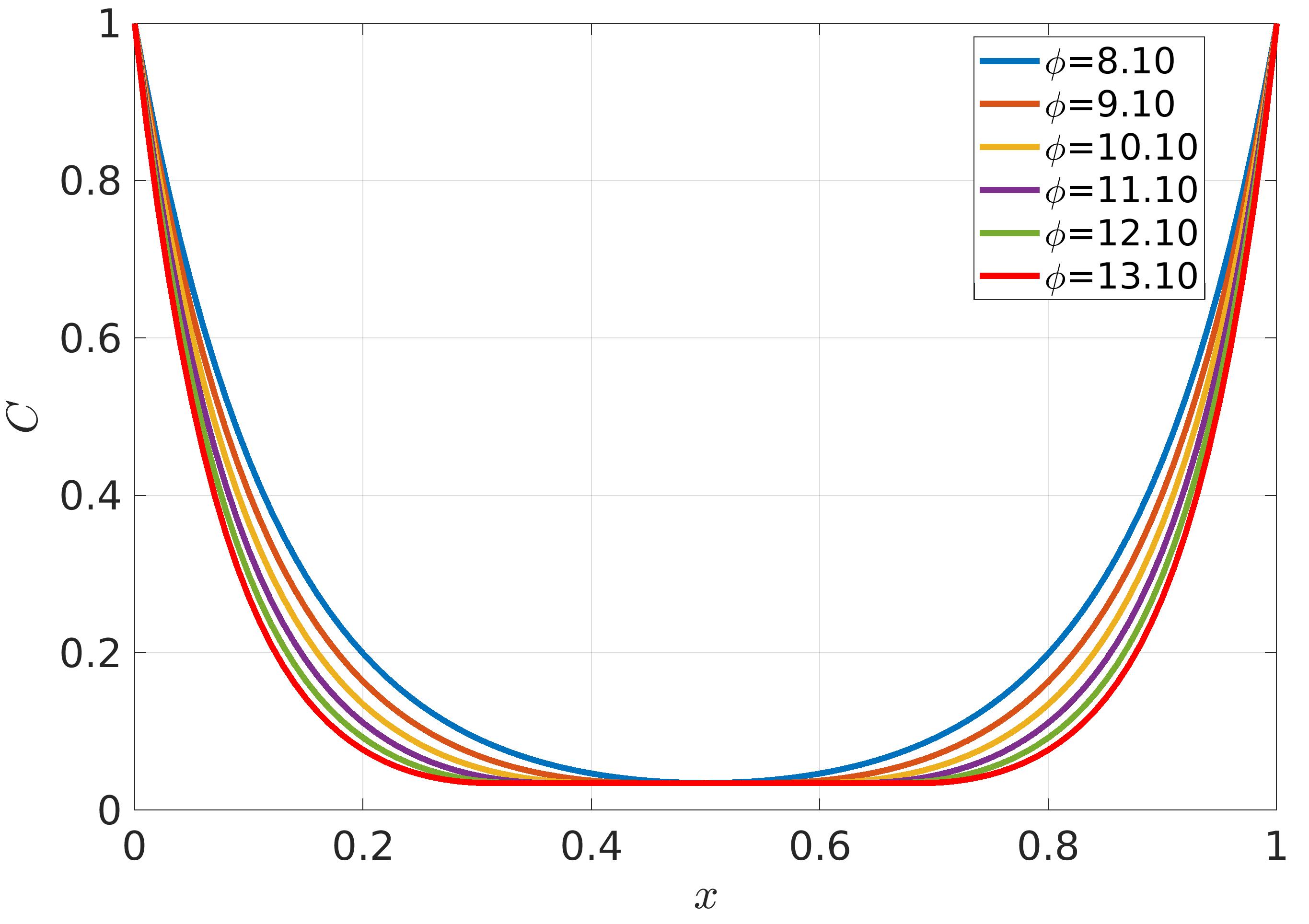}
		\caption{}
		\label{fig:Post_necro_oxygen} 
	\end{subfigure}	
	\begin{subfigure}{0.49\textwidth}
		\centering
		\includegraphics[scale=0.08]{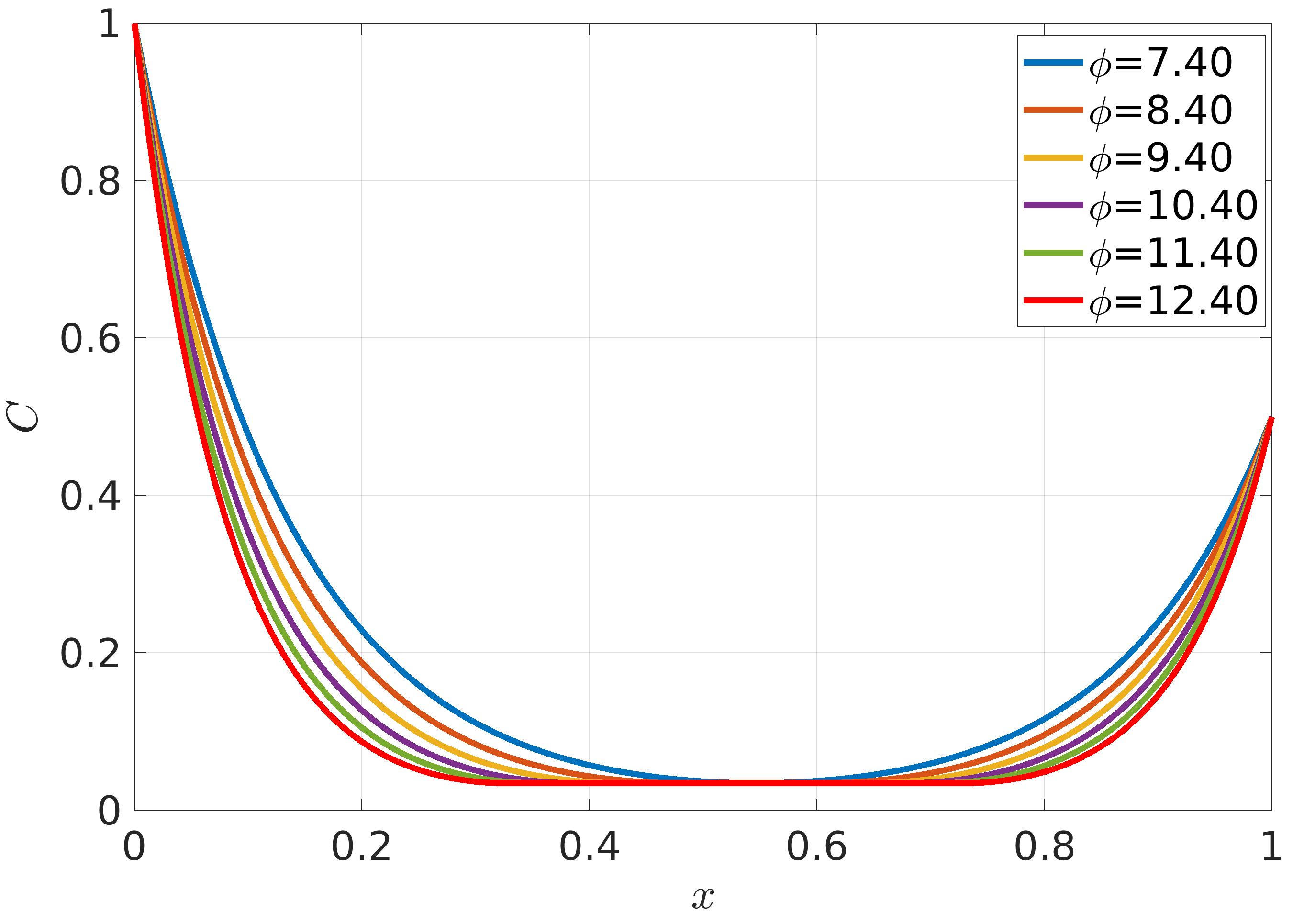}
		\caption{}
		\label{fig:oxygen05_post_necro} 
	\end{subfigure}	
	\begin{subfigure}{0.49\textwidth}
		\centering
		\includegraphics[scale=0.08]{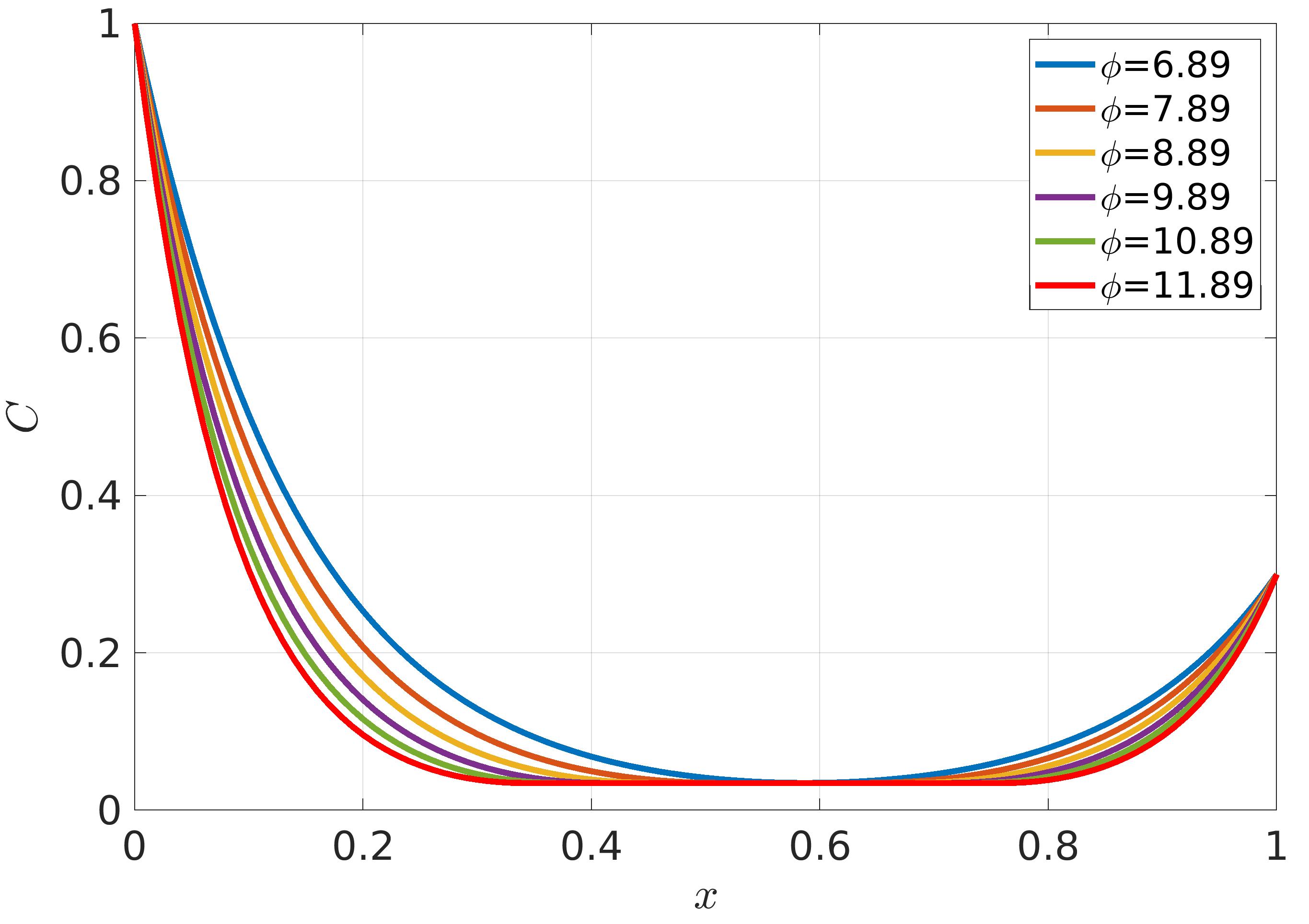}
		\caption{}
		\label{fig:oxygen03_post_necro}
	\end{subfigure}
	\begin{subfigure}{0.49\textwidth}
		\centering
		\includegraphics[scale=0.08]{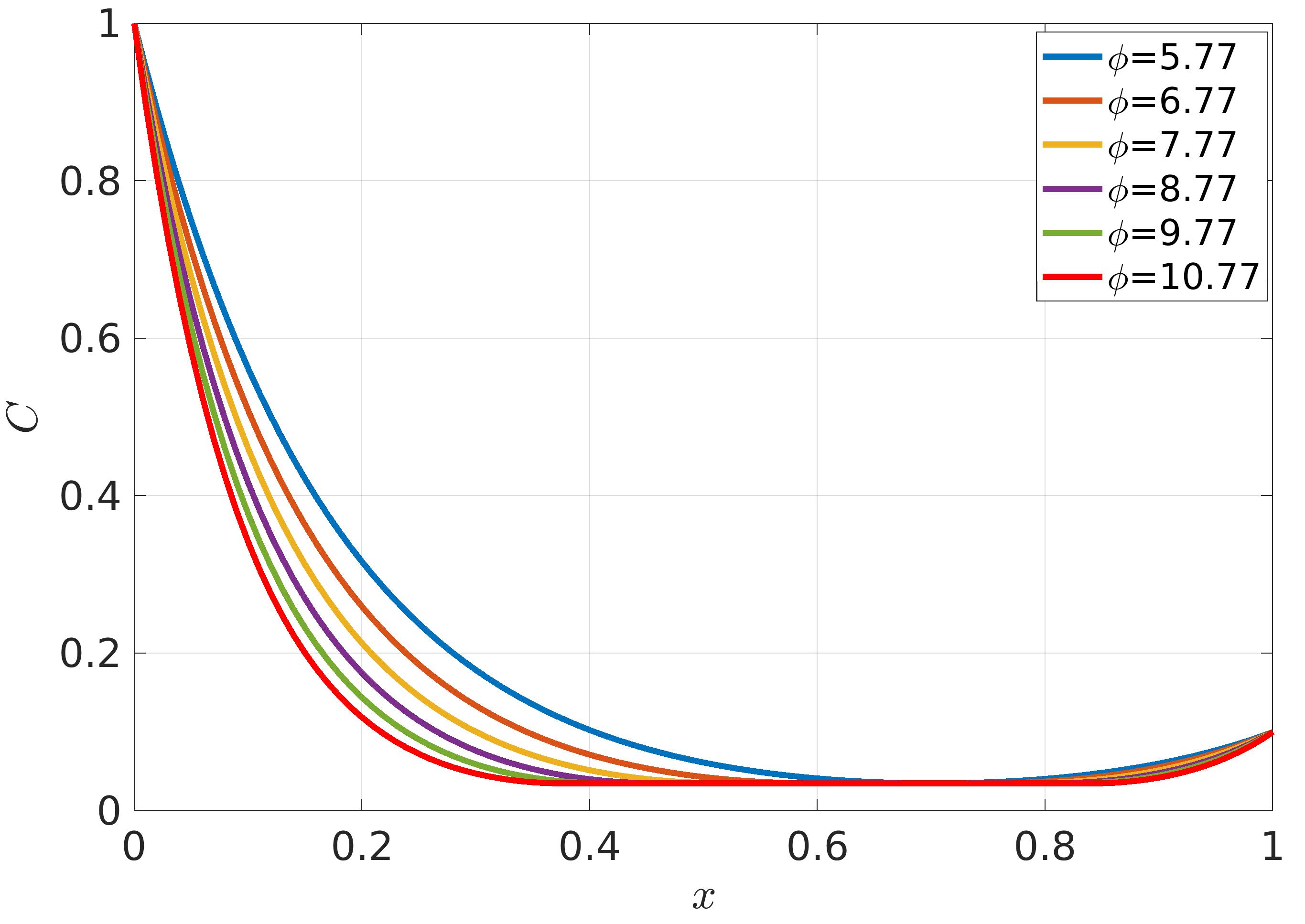}
		\caption{}
		\label{fig:oxygen01_post_necro}
	\end{subfigure}
	
	\caption[Two numerical solutions]{Oxygen concentration profile of tumor tissue having necrotic core for various values of $\phi\; (\ge\phi_{max})$ with (a) $C_l=1$ and $C_r=1$,  (b) $C_l=1$ and $C_r=0.5$, (c) $C_{l}=1$, $C_{r}=0.3$ and (d) $C_{l}=1$, $C_{r}=0.1$.}
	\label{fig:oxygen_tumor_post_necro}
\end{figure}

In this section, our focus is on the effects of oxygen concentration at the tumor boundary on the generation of the necrotic core inside tumor. 
The tumor develops a necrotic core when Thiele Modulus ($\phi$) crosses the value $\phi_{max}$ for the cases ($C_l=C_r=1$; $C_l=1, C_r=0.5$, $C_l=1, C_r=0.3$ and $C_l=1, C_r=0.1$), which are available in Table \ref{tab:Theile number}. Figures \ref{fig:Post_necro_oxygen}-\ref{fig:oxygen01_post_necro} portray the corresponding oxygen profiles. One can notice that the oxygen profile has a central plateau region in the tumor. It occurs because oxygen is not consumed at this plateau region as tumor cells die. The diameter of the necrotic core can be found as $|x_{N,r}-x_{N,l}|$ for the specific value of $\phi\; (>\phi_{max})$, where $x_{N,l}$ and $x_{N,r}$ are the boundary points of the necrotic core, given by the Eqs. \eqref{eq:left_necro} and \eqref{eq:right_necro} respectively. The necrotic radius is calculated for the tumor under different oxygen concentrations at the right boundary to examine the effects of oxygen concentration on necrotic core formation which is shown in Table \ref{tab:Necro_diameter}. It can be concluded that the size of the necrotic core increases with the increase in the ratio of oxygen concentrations at the right and left boundaries (i.e., $\frac{C_r}{C_l}$). Hence, it hints that tumors have larger necrotic cores in an in-vivo setting than that in an in-vitro setting.
\begin{table}[h!]
	\centering
	\caption{Diameter of necrotic core}
	\begin{tabular}{ p{2cm} p{2cm} p{2cm} p{3cm}}
		\hline
		\multicolumn{4} {  c  }{In $C_l=C_r=1$}\\
		\hline
		 $\phi\;(>\phi_{max})$& $x_{N,l}$&$x_{N,r}$&$|x_{N,r}-x_{N,l}|$\\
		\hline
		 $9$ &$0.4501$ &$ 0.5499$ & $0.0998$ \\
		 $10$ &$0.4051$ &$0.5949$&$0.1898$\\
		 $11$ &$0.3683$ &$0.6317$&$0.2635$\\
		 $12$ &$0.3376$ &$0.6624$&$ 0.3248$\\
		\hline
		\multicolumn{4} {  c  }{In $C_l=1\; C_r=0.5$}\\
		\hline
		$\phi\;(>\phi_{max})$& $x_{N,l}$&$x_{N,r}$&$|x_{N,r}-x_{N,l}|$\\
		\hline
		$9$ &$0.4501$ &$0.6270$ & $0.17689$ \\
		$10$ &$0.40509$ &$0.66431$& $0.25921$\\
		$11$ &$0.36827$ &$0.69482$& $0.32655$\\
		$12$ &$0.33758$ &$0.72026$&$0.38267$\\
		\hline
		\multicolumn{4} {  c  }{In $C_l=1\; C_r=0.3$}\\
		\hline
		$\phi\;(>\phi_{max})$& $x_{N,l}$&$x_{N,r}$&$|x_{N,r}-x_{N,l}|$\\
		\hline
		$9$ &$0.45011$ &$0.68401$ & $0.23389$ \\
		$10$ &$0.40509$ &$0.71560$&$0.31051$\\
		$11$ &$0.36827$ &$0.74146$&$0.37319$\\
		$12$ &$0.33758$ &$0.76301$&$0.42542$\\
		\hline
		\multicolumn{4} {  c  }{In $C_l=1\; C_r=0.1$}\\
		\hline
		$\phi\;(>\phi_{max})$& $x_{N,l}$&$x_{N,r}$&$|x_{N,r}-x_{N,l}|$\\
		\hline
		$9$ &$0.45011$ &$0.80923$ & $0.35911$  \\
		$10$ &$0.40509$ &$0.82830$&$0.42321$\\
		$11$ &$0.368271$ &$0.843913$&$0.47564$\\
		$12$ &$0.33758$ &$0.85692$&$0.51933$\\
		\hline
	\end{tabular}
	\label{tab:Necro_diameter}
\end{table}

\subsection{Lactate profile inside the tumor prior to the formation of a necrotic core}
\begin{figure}[h!]
	\begin{subfigure}{0.49\textwidth}
		\centering
		\includegraphics[scale=0.08]{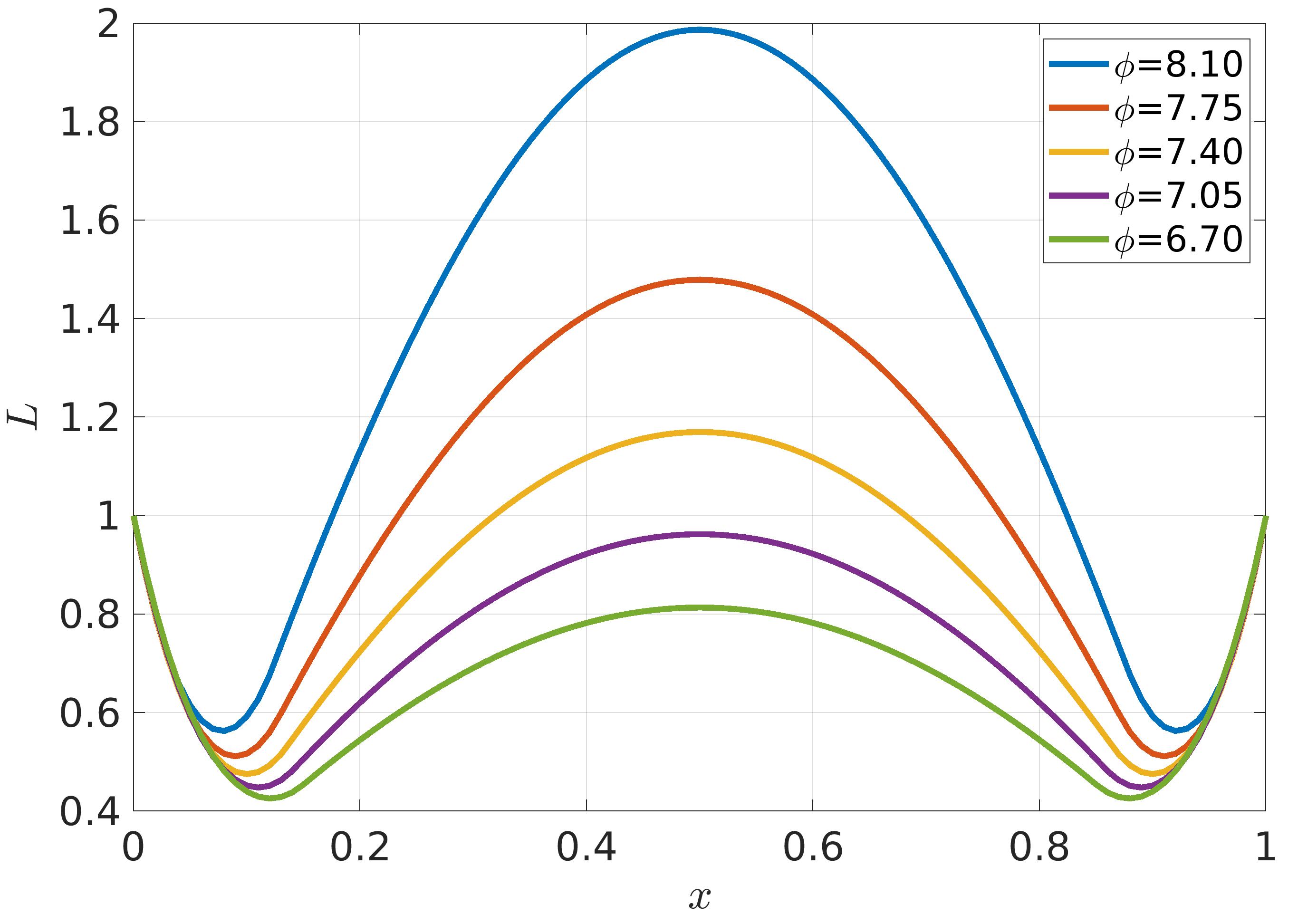}
		\caption{}
		\label{fig:pre_necro_lactate} 
	\end{subfigure}	
	\begin{subfigure}{0.49\textwidth}
		\centering
		\includegraphics[scale=0.08]{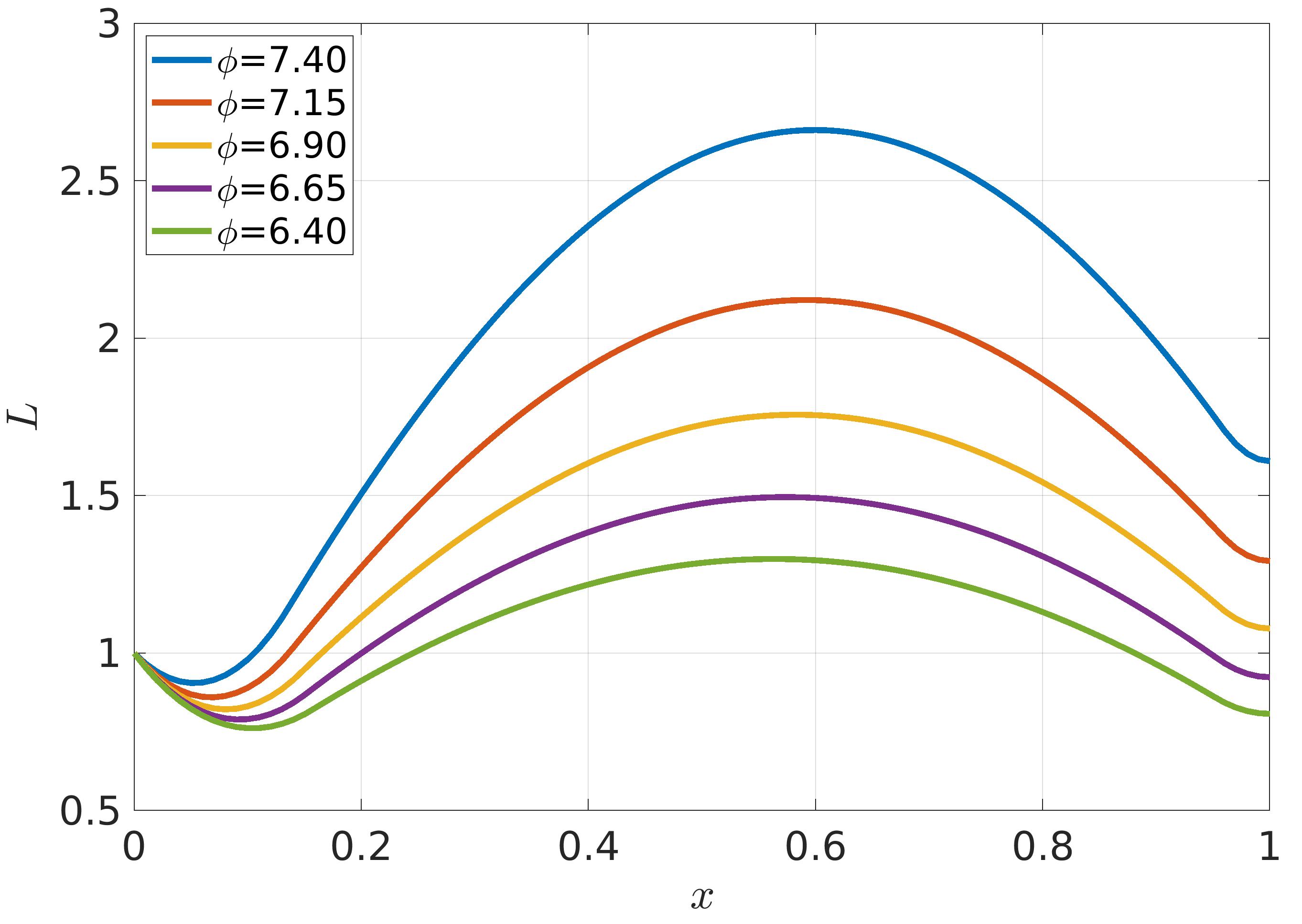}
		\caption{}
		\label{fig:lactate05_pre_necro} 
	\end{subfigure}	
	\begin{subfigure}{0.49\textwidth}
		\centering
		\includegraphics[scale=0.08]{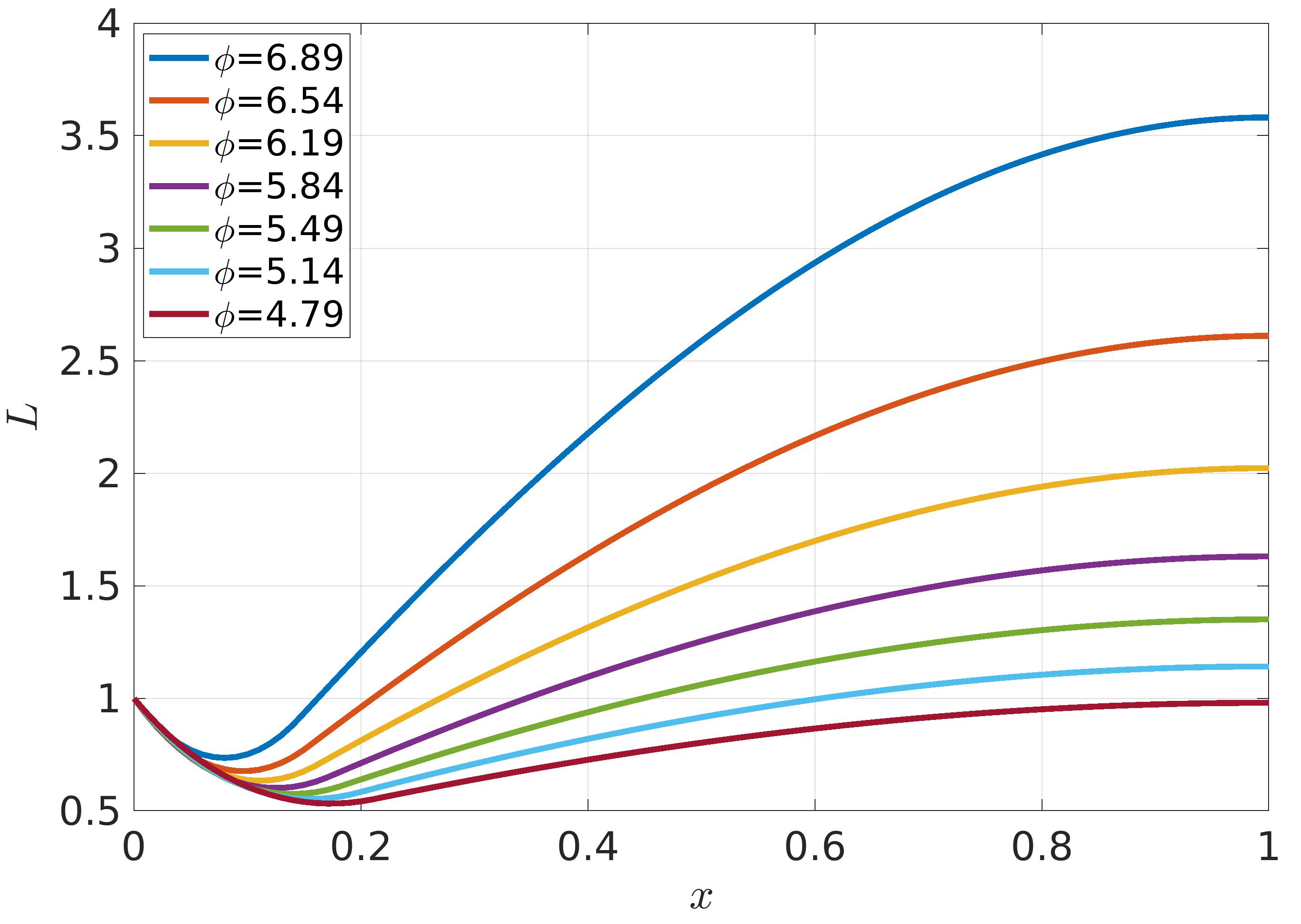}
		\caption{}
		\label{fig:lactate03_pre_necro}
	\end{subfigure}
	\begin{subfigure}{0.49\textwidth}
		\centering
		\includegraphics[scale=0.08]{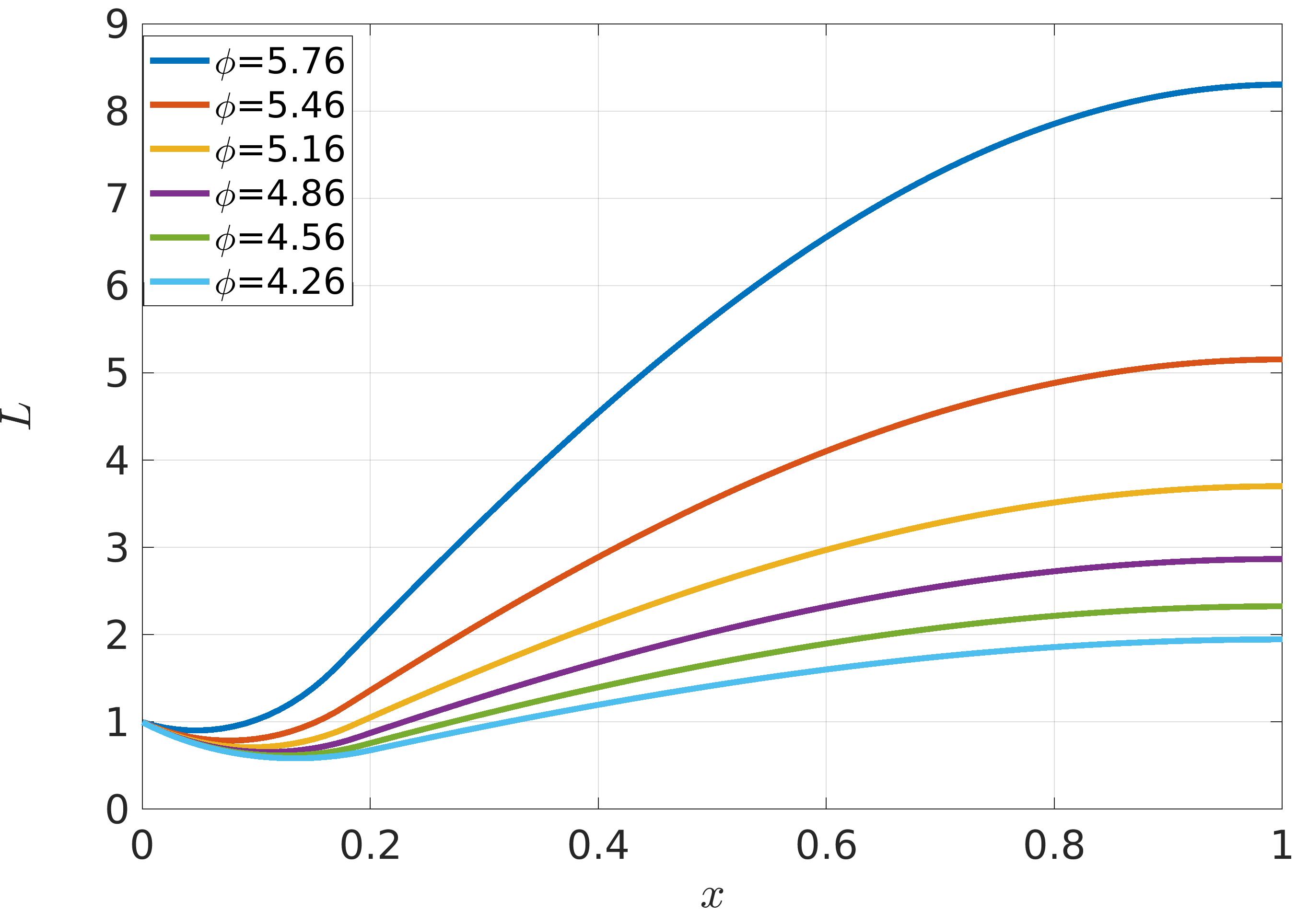}
		\caption{}
		\label{fig:lactate01_pre_necro}
	\end{subfigure}
	
	\caption[Two numerical solutions]{Lactate concentration profile of tumor tissue having necrotic core for various values of $\phi\; (\le\phi_{max})$ with (a) $C_l=1$ and $C_r=1$,  (b) $C_l=1$ and $C_r=0.5$, (c) $C_{l}=1$, $C_{r}=0.3$ and (d) $C_{l}=1$, $C_{r}=0.1$.}
	\label{fig:lactate_tumor_pre_necro}
\end{figure}

This section focuses on the investigation of lactate accumulation in tumors considering the effects of tumor size in situations like prior to the formation of a necrotic core and after the emergence of a hypoxic region. We calculate the minimum $\phi$ value at which tumor begins developing a hypoxic region as per the same evaluation process employed to obtain $\phi_{max}$. By substituting $C_H$ for $C_N$ in Eq. \eqref{eq:transcendental}, we determine the resulting value, denoted as $\phi_{min}$. For the in-vitro case ($C_l=C_r=1$), $\phi_{min}=3.3025$ and for a in-vivo case ($C_l=1$, $C_r=0.5$), $\phi_{min}=2.4667$. Hence, for $\phi_{min}<\phi<\phi_{max}$, tumor has no necrotic region. When the oxygen levels fall below the hypoxic threshold at the right boundary, tumor cells in the vicinity of right boundary no longer have access to a normoxic environment. This leads to the halt of lactate consumption and the onset of a site for lactate production. For the cases ($C_l=1$, $C_r=0.5$ and $C_l=1$, $C_r=0.1$), $\phi_{min}$ is zero. The lactate profile are displayed in Figure \ref{fig:lactate_tumor_pre_necro} for the in-vitro and in-vivo setups. One can observe that lactate is distributed symmetrically about the tumor center and it is consumed at both the ends of the tumor for an in-vitro case (Figure \ref{fig:pre_necro_lactate}). Lactate starts to accumulate at central region with the increase in size of tumor i.e., size of hypoxic region. 
In Figures \ref{fig:lactate05_pre_necro}, \ref{fig:lactate03_pre_necro} and \ref{fig:lactate01_pre_necro}, lactate profiles are portrayed corresponding to in-vivo situations of $C_l=1$, $C_r=0.5$; $C_l=1$, $C_r=0.3$ and $C_l=1$, $C_r=0.1$ respectively. Since, in in-vivo situations, lactate is not flushed out at the end far from blood vessel, lactate accumulation is observed at the boundary having low oxygen concentration compared with other end oxygen concentration. 

From this section, one can conclude that lactate accumulation increases with the increase in oxygen gradient at the boundary.
\subsection{Lactate profile after the formation of a necrotic core in tumor}
\begin{figure}[h!]
	\begin{subfigure}{0.49\textwidth}
		\centering
		\includegraphics[scale=0.08]{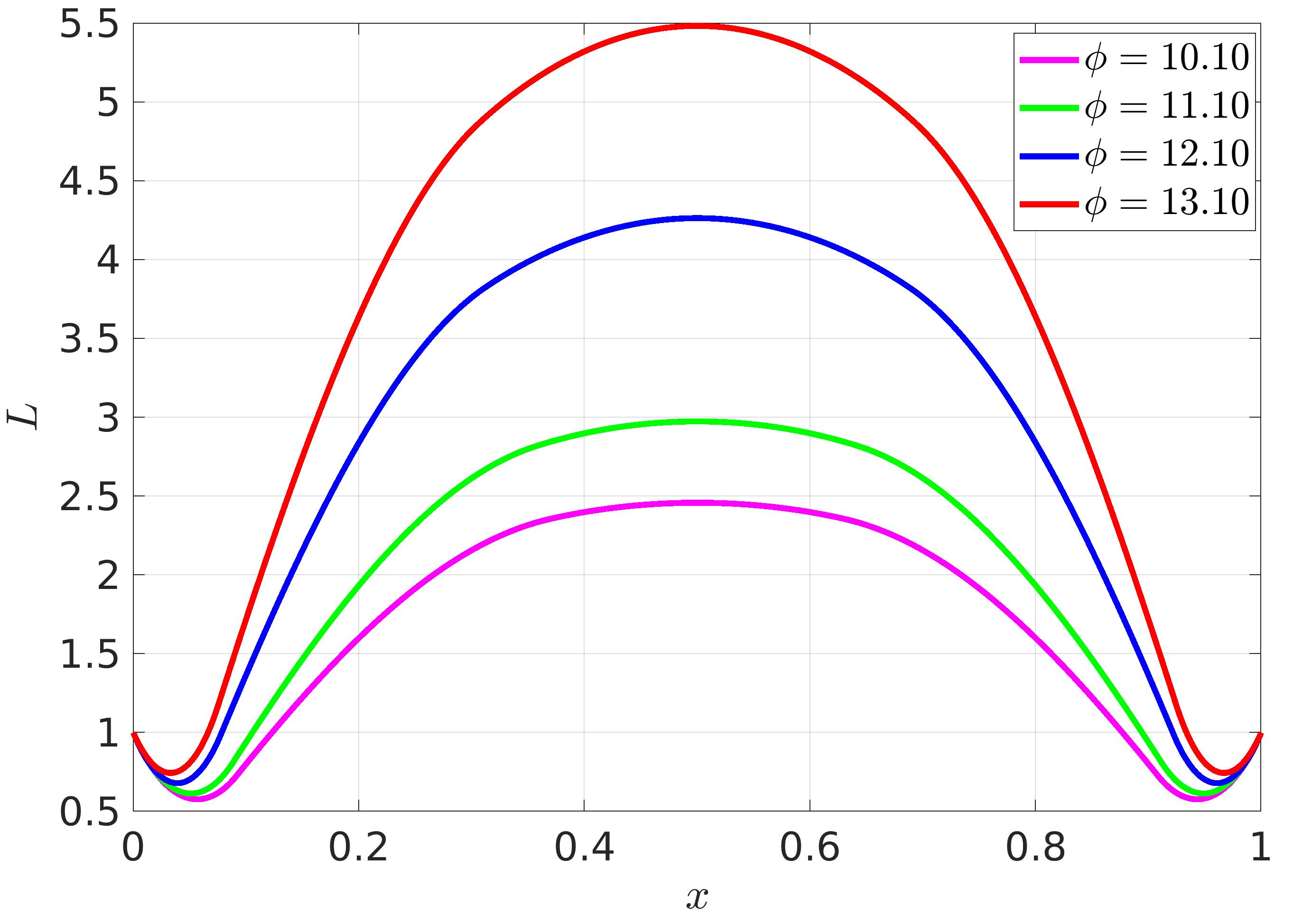}
		\caption{}
		\label{fig:post_necro_lactate} 
	\end{subfigure}	
	\begin{subfigure}{0.49\textwidth}
		\centering
		\includegraphics[scale=0.08]{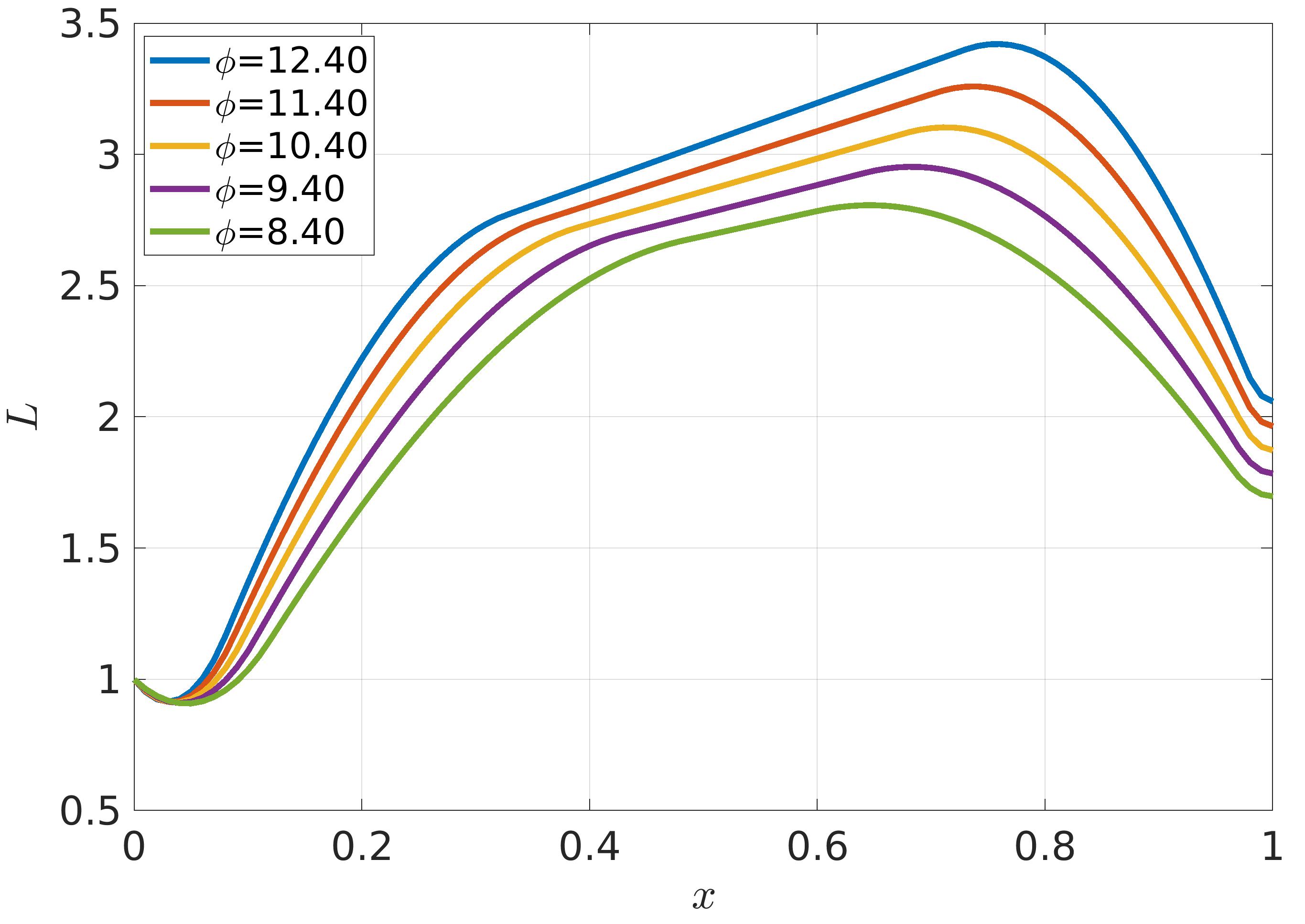}
		\caption{}
		\label{fig:lactate05_post_necro} 
	\end{subfigure}	
	\begin{subfigure}{0.49\textwidth}
		\centering
		\includegraphics[scale=0.08]{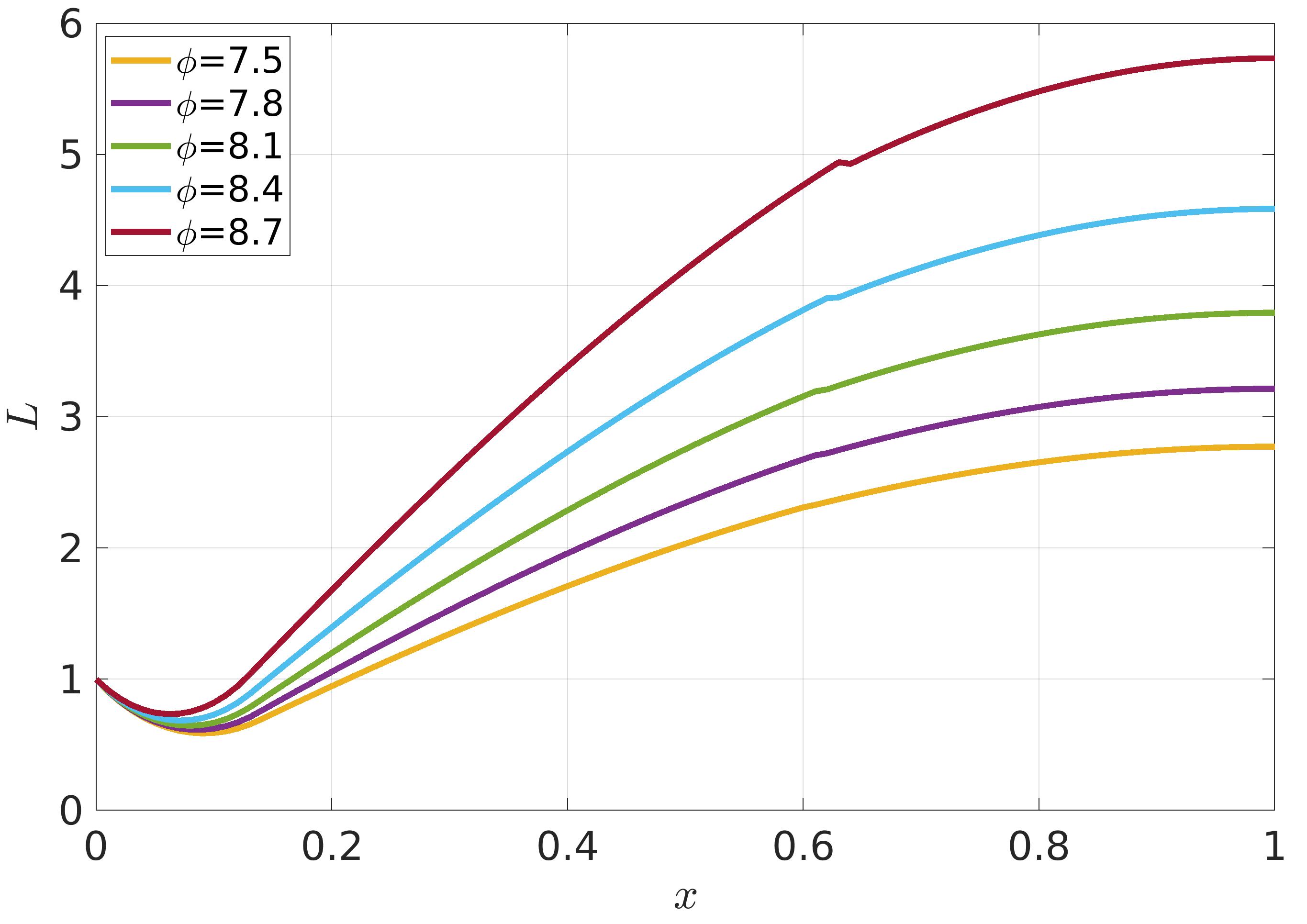}
		\caption{}
		\label{fig:lactate03_post_necro}
	\end{subfigure}
	\begin{subfigure}{0.49\textwidth}
		\centering
		\includegraphics[scale=0.08]{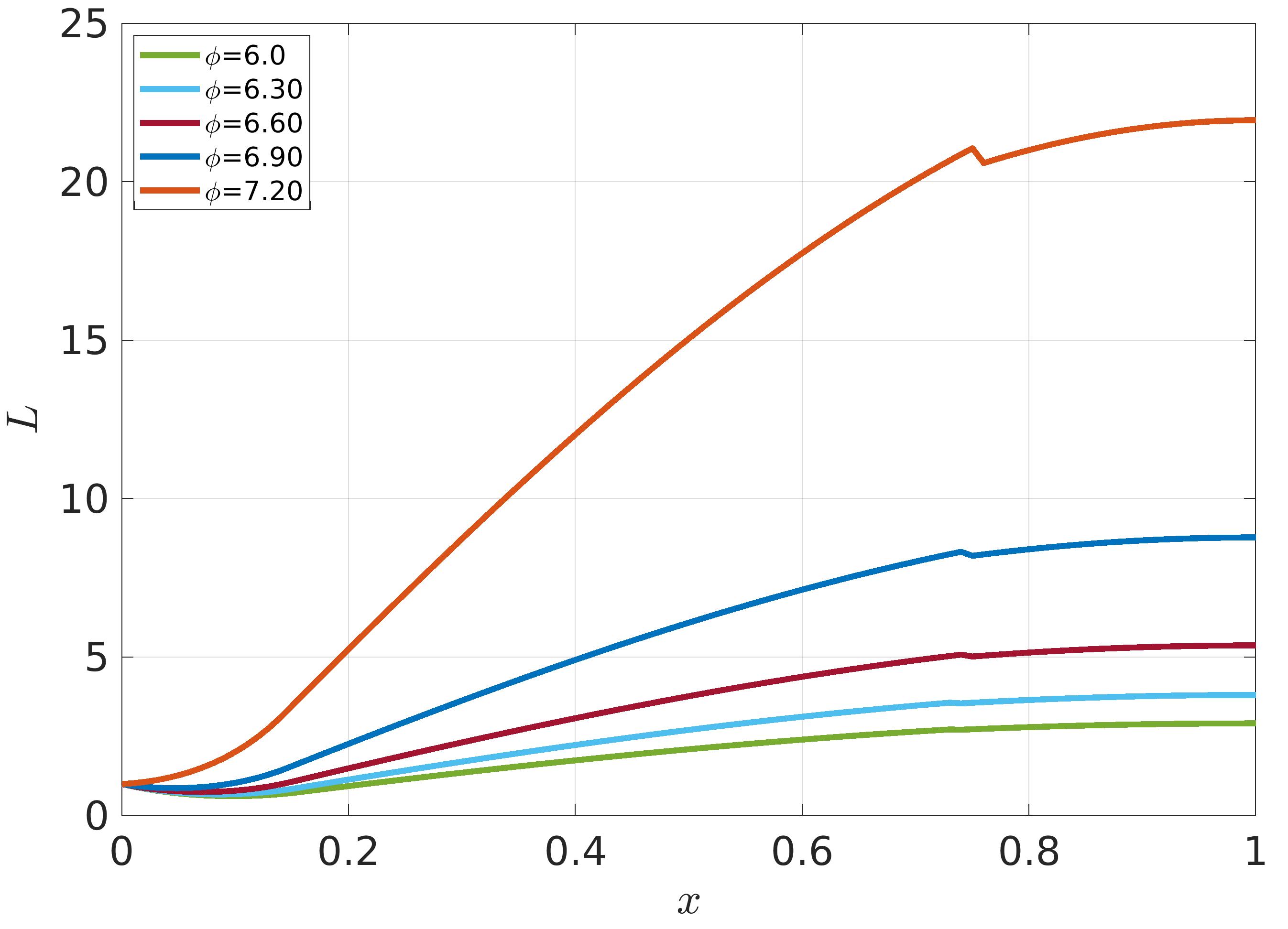}
		\caption{}
		\label{fig:lactate01_post_necro}
	\end{subfigure}
	
	\caption[Two numerical solutions]{Lactate concentration profile of tumor tissue having necrotic core for various values of $\phi\; (\ge\phi_{max})$ with (a) $C_l=1$ and $C_r=1$,  (b) $C_l=1$ and $C_r=0.5$, (c) $C_{l}=1$, $C_{r}=0.3$ and (d) $C_{l}=1$, $C_{r}=0.1$.}
	\label{fig:lactate_tumor_post_necro}
\end{figure}

In this section, we explore the lactate profile inside the tumor in the presence of a necrotic core. The tumor has a necrotic core when $\phi$ surpluses $\phi_{max}$. After the initial formation of the necrotic core within the tumor, there is a noticeable decrease in lactate levels compared to the lactate profile observed when $\phi=\phi_{max}$. This is due to the necrotic region does not produce lactate and lactate gets diffused in the necrotic region. When tumor size increases, hypoxic regions increase and produce enough lactate to refill the lactate level. As a results, the lactate level rises. Lactate profiles are displayed in Figure \ref{fig:lactate_tumor_post_necro}. Interestingly, the impact of lactate consumption on lactate profile appears to be independent of the available lactate concentration in tumors characterized by oxygen concentrations of $C_l=1$ and $C_r=0.1$ at their boundaries (as illustrated in Figure \ref{fig:lactate01_post_necro}). This behavior is likely attributed to the substantial presence of lactate within the tumor microenvironment. But, for in-vitro case ($C_l=C_r=1$), the impact of lactate consumption by normoxic tumor cells on lactate profile inside the tumor is observed (Figure \ref{fig:post_necro_lactate}). Also, for the in-vivo cases ($C_l=1, \; C_r=0.5$ and $C_l=1, \; C_r=0.3$), lactate consumption effects is noticed (Figures \ref{fig:lactate05_post_necro}, \ref{fig:lactate03_post_necro}).

From this section, it can conclude that during the initial stages of necrotic core formation, there is an influence on lactate accumulation within the tumor. As the process advances to later stages, the presence of necrotic core appears to have a diminished impact on the lactate status within the tumor.
\subsection{Analyzing lactate profiles in a non-consumption scenario}	
\begin{figure}[h!]
	\begin{subfigure}{0.49\textwidth}
		\centering
		\includegraphics[scale=0.08]{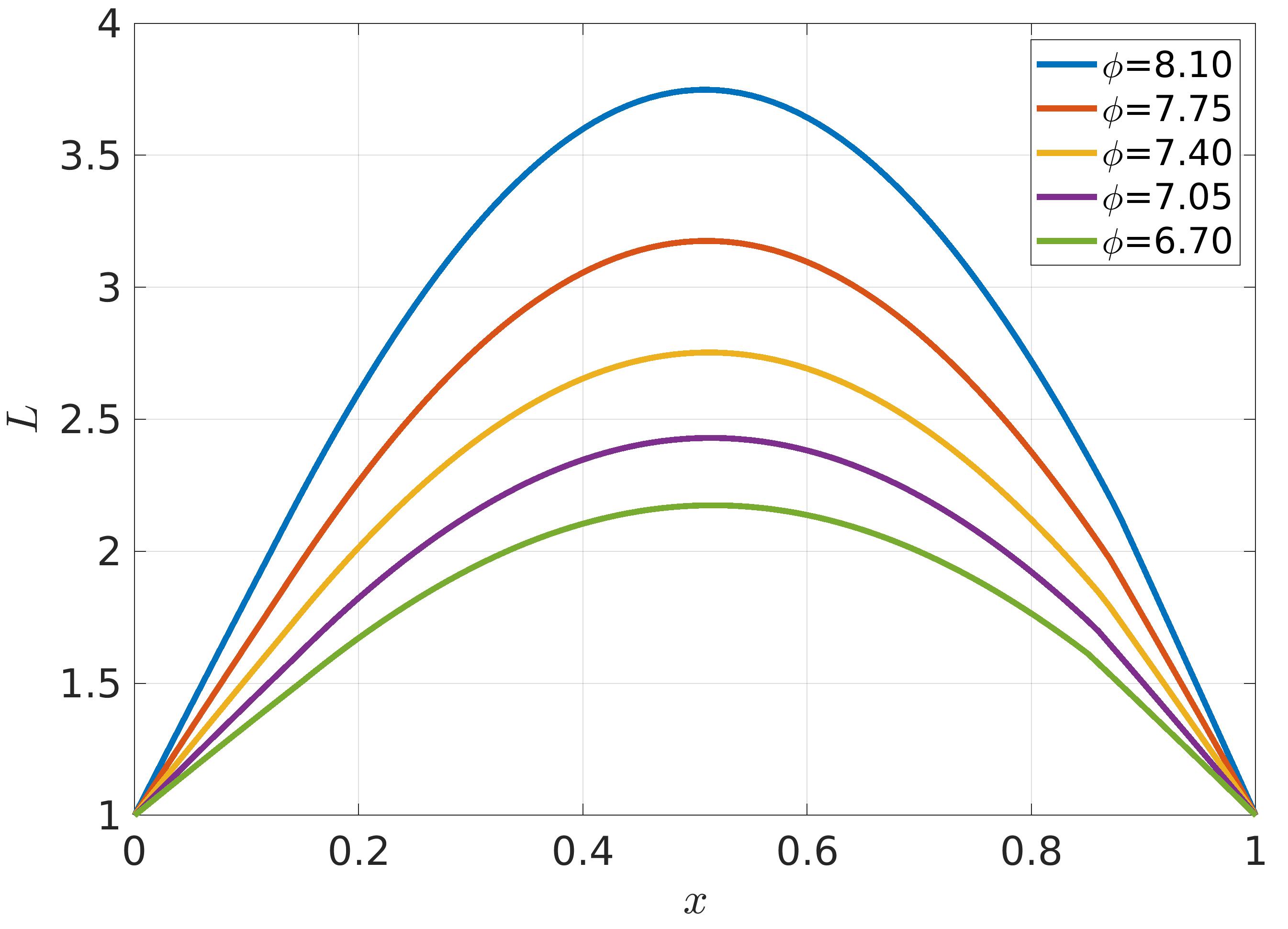}
		\caption{}
		\label{fig:cr1_without_consume} 
	\end{subfigure}	
	\begin{subfigure}{0.49\textwidth}
		\centering
		\includegraphics[scale=0.08]{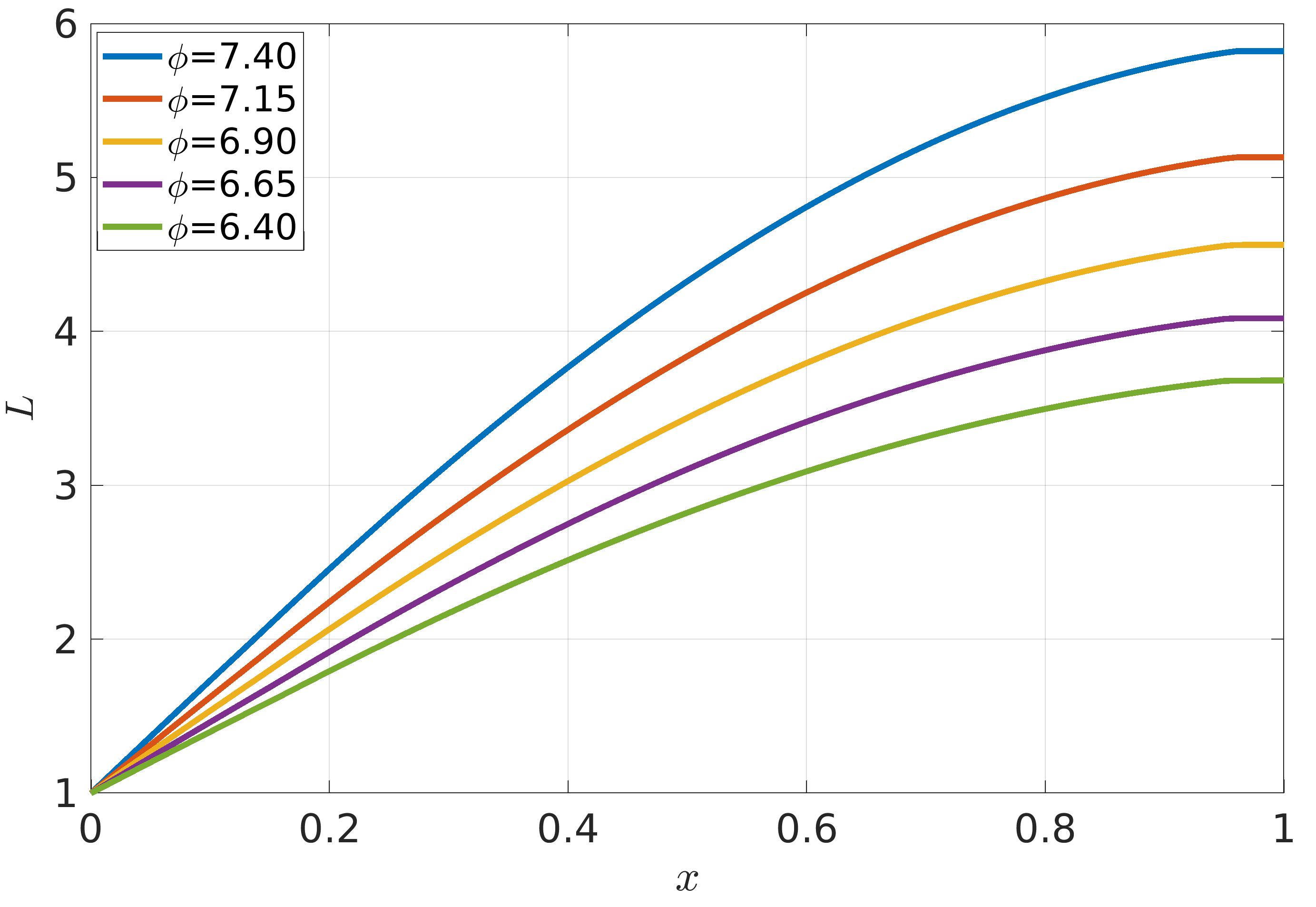}
		\caption{}
		\label{fig:cr0.5_without_consume} 
	\end{subfigure}	
	\begin{subfigure}{0.49\textwidth}
		\centering
		\includegraphics[scale=0.08]{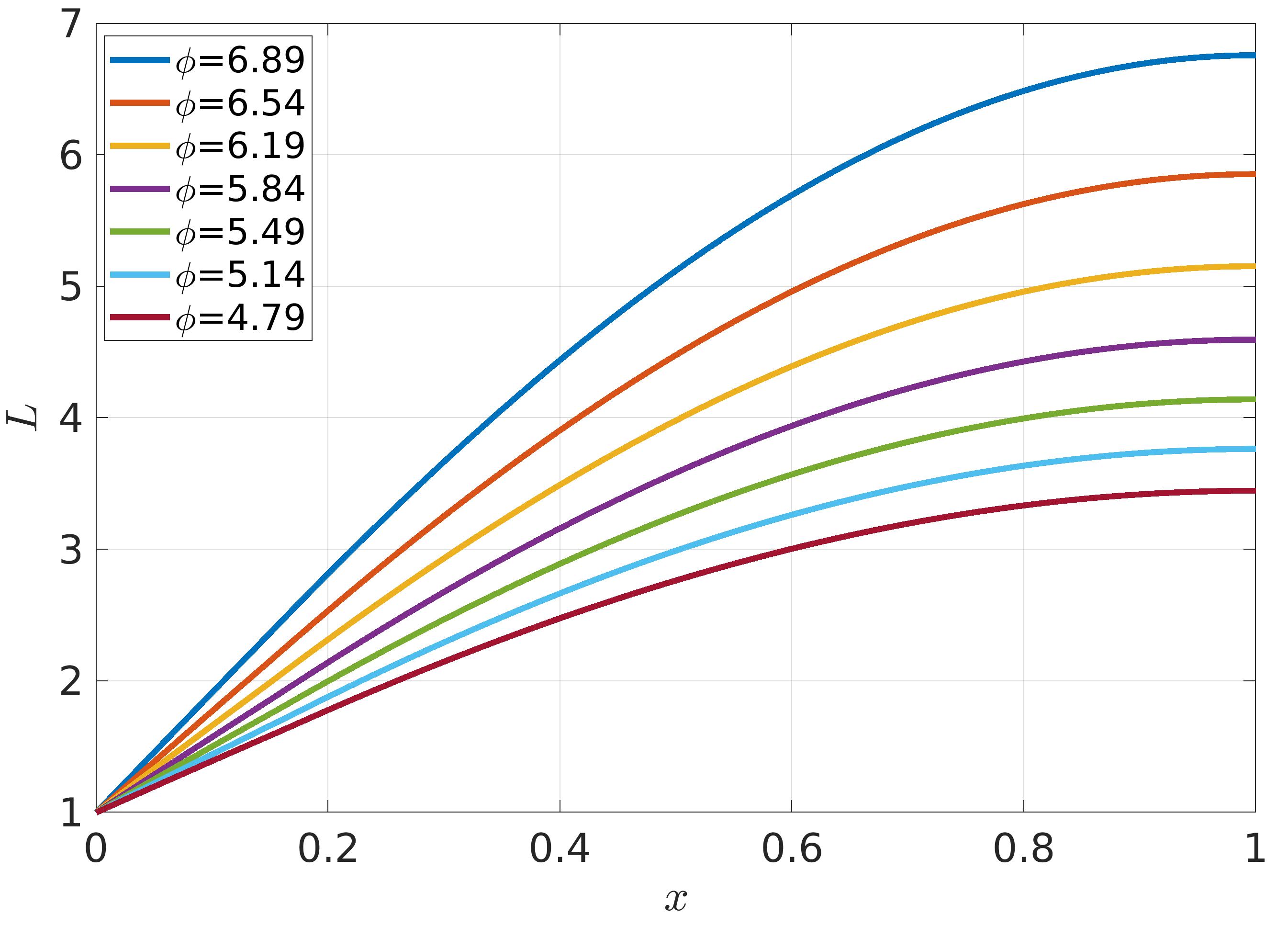}
		\caption{}
		\label{fig:cr0.3_without_consume}
	\end{subfigure}
	\begin{subfigure}{0.49\textwidth}
		\centering
		\includegraphics[scale=0.08]{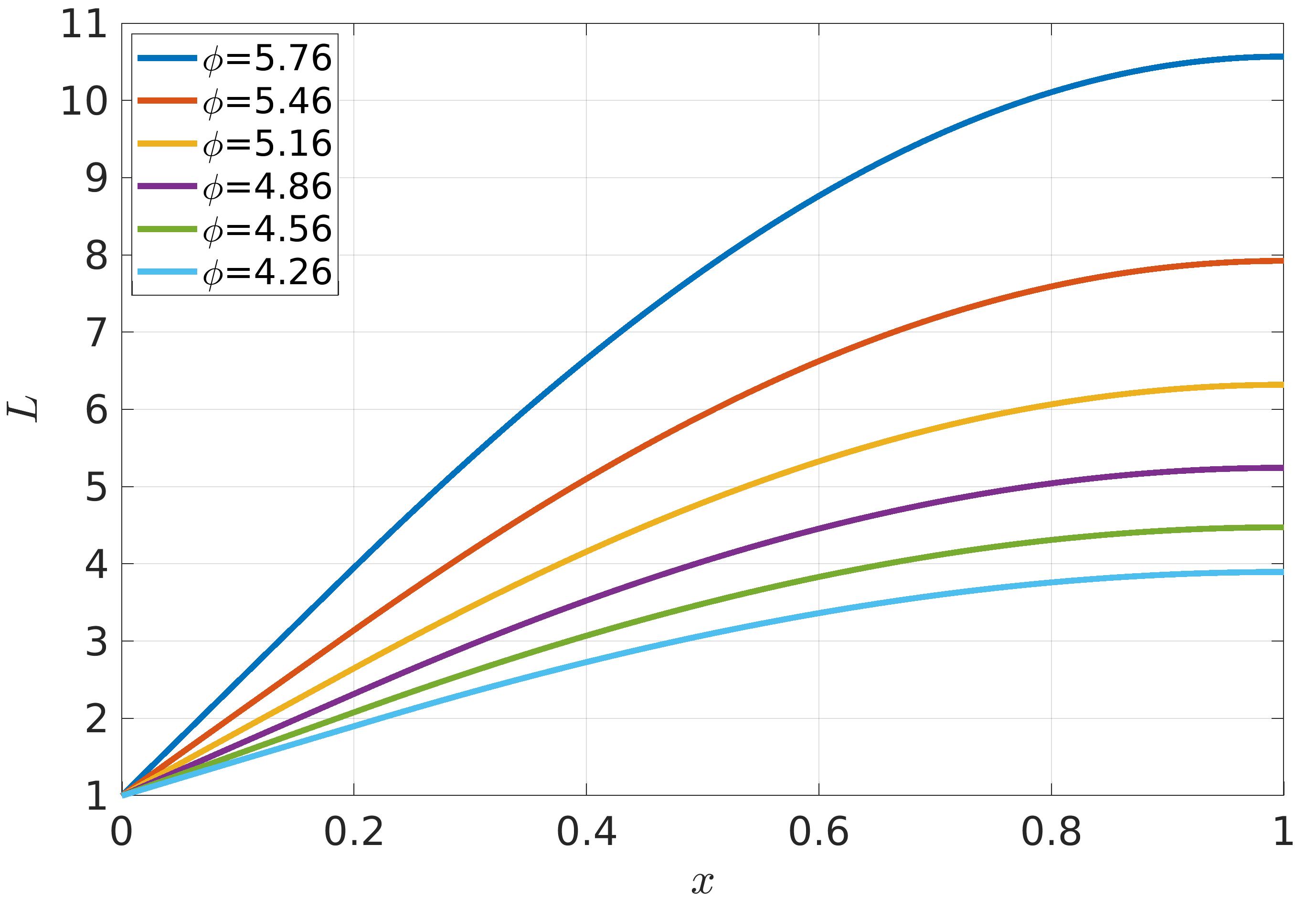}
		\caption{}
		\label{fig:cr0.1_without_consume}
	\end{subfigure}
	
	\caption[Two numerical solutions]{Lactate concentration profile without the consumption in pre-necrotic tumor for various values of $\phi\; (\le\phi_{max})$ with (a) $C_l=1$, $C_r=1$ and $\phi_{max}=8.10$,  (b) $C_l=1$, $C_r=0.5$ and $\phi_{max}=7.40 $, (c) $C_{l}=1$, $C_{r}=0.3$ and $\phi_{max}=6.89 $ and (d) $C_{l}=1$, $C_{r}=0.1$ and $\phi_{max}=5.77$.}
	\label{fig:lactate_without_consume}
\end{figure}

In this section, we study the lactate profile in a tumor when lactate consumption by normoxic tumor cells is not considered. In this context, the governing equation for lactate diffusion and its solutions are presented in \textbf{Appendix} \ref{appendix:a}. The lactate concentration profile for pre-necrotic tumors having oxygen concentration $C_r=1$, $C_r=0.5$, $C_r=0.3$ and $C_r=0.1$ at the right boundary, with $C_l=1$ at the left boundary, are depicted in Figure \ref{fig:lactate_without_consume}. One can see that lactate gets accumulated at the central part of the tumor for an in-vitro tumor (Figure \ref{fig:cr1_without_consume}). In contrast, lactate tends to gather at the boundary having lower level of oxygen concentration for in-vivo cases (Figures \ref{fig:cr0.5_without_consume}, \ref{fig:cr0.3_without_consume}, \ref{fig:cr0.1_without_consume}). Moreover, it can be observed that the lactate level increases with the increase in size of the tumor, irrespective of the growth environment. Also, its level is high compared to that when normoxic tumor cells consume lactate as an alternative metabolic fluid.

One can conclude from this section that lactate accumulation significantly increases when lactate consumption by normoxic tumor cells is not considered. 
To determine the lactate profile of post-necrotic tumors where lactate consumption is absent, we followed the same process in the Appendix \ref{appendix:a}. The predicted results have similar qualitative behavior to the lactate profile in the pre-necrotic tumor (results are not shown).

\section{Discussion}	
In this work, we propose a simplified mathematical model for lactate and oxygen interplay in a tumor in  in-vitro and in-vitro situations. The effects of oxygen concentration  on lactate generation at the tumor boundaries are analyzed. The range of lactate concentration in the normal physiological cord is $0.5-2$ $\mu M$. In contrast, tumor tissue increases lactate concentration up to $40$ $\mu M$ \cite{phypers2006lactate}. Healthy cells produce energy by oxidative phosphorylation, and this process is slow. On the other hand, tumor cells are highly proliferative. Therefore, they need energy in quick time. To meet this need, tumor cells switch their metabolic pathway from oxidative phosphorylation to aerobic glycolysis \cite{warburg1927metabolism}. During aerobic glycolysis, lactate is produced as a byproduct. \citet{sonveaux2008} showed that lactate is not only a metabolism byproduct but also it is consumed by tumors in oxygen-rich areas. Lactate is taking part in energy production. Lactate helps tumors to be more aggressive and creates a growth-friendly environment. Along with this, lactate can change the therapeutic effectiveness and hinder achieving maximal therapy outcomes\cite{rossi2022lactate}. So, analysis of lactate concentration is crucial for the disease status, therapeutic outcome and invasiveness of the disease. In reality, generation of lactate in tumor is a complex chain of network  \cite{lule2013metabolic,venkatasubramanian2006incorporating}. However, the present model incorporates only the oxygen and lactate interplay.

In in-vitro setup, oxygen availability is limitless at the tumor boundary. In this study, we assumed that oxygen concentration remains equal at both the edges in in-vitro situation. The oxygen profile for various sizes is analyzed (Figure \ref{fig:oxygen} and \ref{fig:Post_necro_oxygen}). The oxygen concentration touches the necrotic threshold value for Thiele modulus $\phi_{max}=8.10$ (Figure \ref{fig:oxygen}) and after that tumor develops a necrotic core (Figure \ref{fig:Post_necro_oxygen}). In the oxygen concentration profiles, after the formation of a necrotic core, a plateau region is developed about the center of the tumor. With the increase in the value of $\phi$, this plateau region increases. It arises as tumor cells die and does not consume oxygen. This type of observation is also found in experimental setup \cite{1982muellerl}. However, when tumor is developed in the human body, oxygen concentrations  are not equal at the boundaries. Considering unequal oxygen concentration at the boundaries, it is found that the tumor starts to form necrotic core with the value of $\phi$ with less value of $\phi_{max}$ for an in-vitro setup. The value of $\phi_{max}$ declines with a decline in the value of $C_r$ at the right boundary of the tumor (Table \ref{tab:Theile number}). When $\phi$ exceeds the value of $\phi_{max}$, necrotic region is developed and it is shifted towards the lower oxygen site (Figures \ref{fig:oxygen01_post_necro}, \ref{fig:oxygen03_post_necro}, \ref{fig:oxygen05_post_necro}). The diameter of necrotic core increases as $C_r$ decreases. In any in-vivo case, the necrotic region is not formed at the tumor's center.  The configuration of the necrotic core experiences dynamic changes influenced by the uneven distribution of oxygen levels at its peripheries. This intricate interplays between oxygen concentration results in a corresponding asymmetry not only in the necrotic core but also in the overall tumor morphology. These phenomena closely aligns with the experimental results reported by \citet{1982muellerl}.

 In this study, the lactate equation primarily considers the generation of lactate by hypoxic cells and its consumption by the outer layer of well-oxygenated tumor cells. The equation additionally takes into consideration the production of lactate even in aerobic environments via alternate routes, like amino acid conversion \cite{bossart1979lactate}. The lactate accumulation is affected due to its consumption by normoxic cells (Figures  \ref{fig:lactate_tumor_pre_necro}, \ref{fig:lactate_tumor_post_necro}). In the in-vitro case, lactate accumulation is not very high compared to that in in-vivo cases. This phenomenon can be attributed to two underlying facts. Firstly, in the in-vitro setup, lactate is flushed out through both the tumor boundaries. Conversely, in an in-vivo scenario, lactate is effectively flushed out only from the boundary of the tumor that is closer to the blood vessel. Secondly, in the in-vitro case, the tumor develops normoxic regions of equal length at both the boundaries. We observe the effects of the necrotic zone on lactate accumulation during the early stages of the creation of the necrotic core. During this phase, irrespective of the growth environment, there is a decrease in lactate levels.  It may occur because our model directly takes into account the fact that hypoxic tumor cells are the primary source of lactate production.  As the necrotic zone initiates formation, lactate diffuses towards the newly developed necrotic region. However, the lactate level increases with the size increment of the tumor. It may happen as hypoxic regions also increase with the increase in tumor size. When the oxygen concentration at the right boundary falls below the hypoxic critical value, the tumor tissue no longer will have any normoxic area on that side. As a result, the area near the right boundary of the tumor is converted into a hypoxic region and acts as a mine of lactate. So, the lactate level becomes higher in the cases $C_r=0.3$, $C_r=0.1$ compared to that for the cases $C_r=0.5$, $C_r=1$ (Figures  \ref{fig:post_necro_lactate}, \ref{fig:lactate05_post_necro}, \ref{fig:lactate03_post_necro}, \ref{fig:lactate01_post_necro}). It suggests that the lactate level in an in-vitro tumor is lower than that in an in-vivo tumor.  Also, one can notice that lactate level does not surpass the experimentally observed value of $40\; \mu M$ for $C_r=1,\;C_r=0.5\;\text{ and }\;C_r=0.3$ but its value crosses the expected value for the case $C_r=0.1$. 
 

The overall analysis indicates that oxygen concentrations at the tumor boundaries have notable impacts on lactate accumulation . 
	
\section{Conclusions}
In this article, we proposed a mathematical model to explore the lactate dynamics inside the tumor tissue. For the first time, lactate that serves as an alternative fuel source in oxygen-rich environments is incorporated, encompassing both in in-vivo and in-vitro scenarios within a single mathematical model. The model is solved analytically. The diameter of necrotic core is also calculated analytically. The following conclusive remarks are drawn from the presents study.
\begin{enumerate}
	\item The size of necrotic core of a tumor in the in-vivo scenario is smaller compared to that in the in-vitro setup.
	\item If the oxygen concentration at one boundary is less than that at the other boundary, it prompts the necrotic core to reposition itself towards the boundary with the lower oxygen concentration.  
	\item In in-vivo environments, lactate accumulation constitutes a larger fraction compared to that in an in-vitro setting.
	\item  In an in-vitro setup, lactate accumulation is predominantly found at the tumor core, while in an in-vivo environment, it shifts towards the boundary with lower oxygen concentration.
	\item A sharp elevation in lactate concentration occurs at the right boundary the oxygen concentration falls below the hypoxic threshold.
	\item The effects of  presence of the necrotic core on lactate concentration  is seen when the necrotic core starts to form. However, at a later stage, no significant impact is observed.  
\end{enumerate}

The acidic environmental of the tumor is a biomarker of detection of tumor tissue. Hence, tumor acidosis may help to navigate the disease site in the body.

\section*{Acknowledgments}
The first author of this article thanks to the Ministry of Education, Govt. of India for research fellowship and the Indian Institute of Technology Guwahati, India for the support provided during the period of this work.

\section*{Appendix}\label{appendix:a}
When tumor does not consume lactate (i.e, $\beta=0$ in Eq. \eqref{eq:non_dim_lactate}), then the lactate reaction diffusion equation takes the form for pre-necrotic and post-necrotic tumor as follows
 \begin{equation}\label{eq:lactate_pre_necro_without_consume}
	\left.\begin{aligned}
		0&=\frac{d^2 L}{d x^2}+d\phi^2 \;\;\text{for}\;\; x\in[0,x_{H,l}]\cup[x_{H,r},1],\\
		0&=\frac{d^2 L}{d x^2}+\phi^2\alpha L+d\phi^2 \;\;\text{for}\;\; x\in[x_{H,l},x_{H,r}],
	\end{aligned}\right\}\text{for pre-necrotic tumor}
\end{equation}
and 
\begin{equation}\label{eq:lac_full_tumor_without_consume}
	\left.\begin{aligned}
		0&=\frac{d^2 L}{d x^2}+d\phi^2 \;\;\text{for}\;\; x\in[0,x_{H,l}]\cup[x_{H,r},1],\\
		0&=\frac{d^2 L}{d x^2}+\phi^2\alpha L+d \phi^2 \;\;\text{for}\;\; x\in[x_{H,l},x_{N,l}]\cup[x_{N,r},x_{H,r}],\\
		0&=\frac{d^2 L}{d x^2} \;\; \text{for}\;\; x\in [x_{N,l},x_{N,r}].
	\end{aligned}\right\} \text{for post-necrotic tumor}
\end{equation}
Its corresponding solutions are given as,
\begin{equation}\label{eq:solu_lactate_without_consume}
	L(x)= 
	\begin{cases}
		a_{1}x+a_2-\frac{d\phi^2 x^2}{2}\;\; &\text{for}\;\;x\in[0,x_{H,l}],\\
		a_{3}\cos(\phi\sqrt{\alpha}x)+a_4\sin(\phi\sqrt{\alpha}x)-\frac{d}{\alpha}\;\; &\text{for}\;\; x\in[x_{H,l},x_{H,r}],\\
		a_{5}x+a_{6}-\frac{d\phi^2 x^2}{2}\;\; &\text{for}\;\;x\in[x_{H,r},1].
	\end{cases} 
\end{equation} 
and 
\begin{equation}\label{eq:solu_lactate_post_without_consume}
	L(x)= 
	\begin{cases}
		a_{1}x+a_2-\frac{d\phi^2 x^2}{2},\;\; &\text{for}\;\;x\in[0,x_{H,l}],\\
		a_{3}\cos(\phi\sqrt{\alpha}x)+a_4\sin(\phi\sqrt{\alpha}x)-\frac{d}{\alpha},\;\; &\text{for}\;\; x\in[x_{H,l},x_{N,l}]\\
		a_5x+a_6,\;\; &\text{for}\;\; x\in [x_{N,l},x_{N,r}]\\
		a_{7}\cos(\phi\sqrt{\alpha}x)+a_8\sin(\phi\sqrt{\alpha}x)-\frac{d}{\alpha},\;\; &\text{for}\;\; x\in[x_{N,r},x_{H,r}]\\
		a_{9}x+a_{10}-\frac{d\phi^2 x^2}{2},\;\; &\text{for}\;\;x\in[x_{H,r},1].
	\end{cases}
\end{equation}
The coefficients $a_i$'s are determined using the boundary conditions at the boundary points (i.e., $x=0, 1$); and $L$ and its flux are continuous at the interface points (i.e., $x_{H,l}, x_{H,r}, x_{N,l}, x_{N,r}$).
	
	\bibliography{phy_oxy}
	\bibliographystyle{elsarticle-num}

\end{document}